\documentclass[11pt,draftclsnofoot,journal,onecolumn]{IEEEtran}

\IEEEoverridecommandlockouts

\usepackage[dvips]{graphicx}
\usepackage{subfigure}
\usepackage{amsmath}
\usepackage{amssymb}
\usepackage{verbatim}
\usepackage{url}
\usepackage{cite}
\usepackage{cases}
\usepackage{multirow}

\begin{document}

\title{Throughput and Collision Analysis of Multi-Channel Multi-Stage Spectrum Sensing Algorithms}
\author{Wesam Gabran, Przemys{\l}aw Pawe{\l}czak, and Danijela \v{C}abri{\'c}%
\thanks{The authors are with the Department of Electrical Engineering, University of California, Los 
Angeles, 56-125B Engineering IV Building, Los Angeles, CA 90095-1594, USA (email: \{wgabran, przemek, danijela\}@ee.ucla.edu).}
\thanks{Part of this work has been accepted to the proceedings of IEEE DySPAN, Apr. 3--6, 2011, Aachen, Germany, EU~\cite{gabran_submitted_2010}.}}

\maketitle

\begin{abstract}
Multi-stage sensing is a novel concept that refers to a general class of spectrum sensing algorithms that divide the sensing process into a number of sequential stages. The number of sensing stages and the sensing technique per stage can be used to optimize performance with respect to secondary user throughput and the collision probability between primary and secondary users. So far, the impact of multi-stage sensing on network throughput and collision probability for a realistic network model is relatively unexplored. Therefore, we present the first analytical framework which enables performance evaluation of different multi-channel multi-stage spectrum sensing algorithms for Opportunistic Spectrum Access networks. The contribution of our work lies in studying the effect of the following parameters on performance: number of sensing stages, physical layer sensing techniques and durations per each stage, single and parallel channel sensing and access, number of available channels, primary and secondary user traffic, buffering of incoming secondary user traffic, as well as MAC layer sensing algorithms. Analyzed performance metrics include the average secondary user throughput and the average collision probability between primary and secondary users. Our results show that when the probability of primary user mis-detection is constrained, the performance of multi-stage sensing is, in most cases, superior to the single stage sensing counterpart. Besides, prolonged channel observation at the first stage of sensing decreases the collision probability considerably, while keeping the throughput at an acceptable level. Finally, in realistic primary user traffic scenarios, using two stages of sensing provides a good balance between secondary users throughput and collision probability while meeting successful detection constraints subjected by Opportunistic Spectrum Access communication.
\end{abstract}

\section{Introduction}
\label{sec:introduction}

Opportunistic Spectrum Access (OSA) radios are wireless communication devices that exploit temporarily vacant licensed frequency bands ~\cite{staple_spectrum_2004,Zhao_sigprocmag_2007}. To be able to use these spectral vacancies, OSA radios need to sense the activity of the licensed user in the frequency bands of interest. As spectrum sensing is the enabling technology for OSA, many sensing techniques have been proposed in the literature covering a variety of design aspects~\cite{yucek_commsurv_2009}. This includes different sensing mechanisms at the physical layer as well as various collaborative sensing techniques and measurement fusion methods at the link layer.

In a conventional spectrum sensing algorithm, a OSA radio (denoted later as Secondary User (SU) radio) has to periodically scan the licensed spectrum for any incoming Primary (licensed) Users (PUs), and detect the PUs with a predetermined probability. When the SU radio's spectrum sensing physical layer triggers an alarm that a PU exists, i.e. due to a successful detection or a false alarm, the SU radio should stop transmission on the respective channel. However, the sensing procedure might result in a false alarm, thus, reducing the channel access opportunity for the SU radio. The probability of a false alarm can be decreased by, e.g. increasing the sensing time, however, this might lead to a decrease in the throughput achieved by the SU device, as demonstrated in e.g.~\cite{liang_twc_2008,peh_tvt_2009,pawelczak_tvt_2009}. Besides, the probability of false alarm can be decreased by altering the decision threshold regarding the presence of the primary user. However, this leads to an increase in the probability of a collision between the SU radios and PU devices. This in turn may lead to a decease in throughput caused by packet losses at the SU radio.

One of the ways to increase channel utilization in OSA communication is to use multi-stage sensing. In multi-stage sensing, an alarm (due to successful detection or false alarm) on the PU presence is followed by $S-1$ cycles of sensing (and potential transmission). Thus, the presence of a PU would be declared, and the SU radio would vacate a channel, if and only if $S$ consecutive alarms are generated. The sensing technique used at each of the $S$ stages can be independent from one another. Accordingly, multi-stage sensing provides additional degrees of freedom to optimize performance metrics, namely, throughput of SU radio and the collision probability between PUs and SUs.

\subsection{Related Work}

Multi-stage sensing has been introduced formally in the IEEE 802.22 standard~\cite{ieee80222,stevenson_commag09}, with $S=2$, while the first analyses of a multi-stage sensing algorithm have been performed in~\cite{jeon_twc_2008,luo_twc_2009} (with $S=2$ in case of~\cite{luo_twc_2009}) only for a single PU channel. Note that~\cite{jeon_twc_2008,luo_twc_2009} did not consider the random nature of the OSA traffic which has a notable effect on the performance of the OSA network. Authors in~\cite{jeon_tvt_2010} extended their previous work~\cite{jeon_twc_2008} to the case of two physical layer sensing techniques: (i) energy detection and (ii) feature (cyclostationary) detection. Furthermore, perfect detection at the last stage of sensing and zero delay in channel switching is not assumed. In their model, the frequency of sensing the channel increases when the sensing procedure indicates that PU is possibly present. Finally, in~\cite{Park_arxiv_2010} a multi-channel sensing technique with $S=2$ and one sensing algorithm was considered, where each sensing phase occurred consecutively in the same time slot\footnote{Note that the idea of multi-stage sensing is conceptually similar to the multi-dwell algorithms used in CDMA code acquisition analyzed in~\cite[Sec. II]{polydoros_tcom_1984} or in UWB phase tracking analyzed in~\cite[Sec. II]{suwansantisuk_milcom_2008},~\cite[Sec. II]{suwansantisuk_izs_2006}. However, there are two main differences between multi-stage sensing and multi-dwell acquisition that require a novel modeling approach for multi-stage spectrum sensing. Firstly, in the multi-dwell algorithm, sensing is performed at each stage in parallel, i.e. multiple components sense the same signal at the same time. This is contrary to multi-stage spectrum sensing where parallel detection is problematic due to the hardware cost. Secondly, the detected signal for multi-dwell acquisition is stationary in the time domain, which cannot be guaranteed in OSA systems, due to the traffic fluctuation of the PU.}.

\subsection{Our Contribution}

We conclude that the literature lacks a unified analytical framework that evaluates multi-stage spectrum sensing options when $S\geq2$, considering wide range of algorithms and parameter setups, sensing techniques, various traffic conditions for SU and PU radios, SU buffer availability and multi-channel access. In this paper we provide a comprehensive analytical framework that models the multi-stage multi-channel spectrum sensing concept.

Our work is specifically related to the recent work of~\cite{jeon_tvt_2010}, but should be treated as a new approach for analyzing multi-stage spectrum sensing algorithms. In contrary to~\cite{jeon_tvt_2010}, we provide a more realistic model for PU activity, where we consider the PU traffic, not only in the channel where the SU is operating, but also in all available channels. Furthermore, we account for the scarcity in the bandwidth available for OSA operation where our model accounts for a limited pool of available PU channels. Moreover, our model applies to OSA devices that have a single radio operating on a narrowband channel and OSA devices that are defined to have a stack of narrowband radios that can utilize or sense multiple PU channels at the same time. In contrast to~\cite{jeon_tvt_2010}, our model accounts for different traffic models for the OSA network, including constant and variable bit rate streams. Besides, the model covers unbuffered and buffered OSA systems. Performance is quantified by throughput and a new metric, that is not considered in~\cite{jeon_tvt_2010}, which is the collision probability between the PU traffic and SU devices. We use the proposed analytical framework to evaluate three new multi-stage sensing algorithms that aim at optimizing the aforementioned metrics. In the context of these algorithms, we introduce a new sensing mode, denoted as pre-sensing, which targets minimizing the collision probability. Given that~\cite{jeon_tvt_2010} already evaluated the effect of using different physical sensing methods, and channel switching delay, we do not include it in our study. Since our introduced model accounts for the randomness of the PU and SU traffic, and the most widely used method of analyzing network traffic is via Markov chain analysis~\cite[Ch. 11]{vanmieghem_book_2006}, we follow the same approach here, just like~\cite{jeon_tvt_2010,jeon_twc_2008}.

The rest of the paper is organized as follows. The system model is introduced in Section~\ref{sec:system_model}. The analytical model is presented in Section~\ref{sec:analysis}, with the multi-stage sensing algorithm design examples analyzed in Section~\ref{sec:design_examples}. The numerical results are presented in Section~\ref{sec:numerical_results}. Finally, the paper is concluded in Section~\ref{sec:conclusions}.

\section{System Model}
\label{sec:system_model}

\subsection{General Assumptions}
\label{sec:general_assumptions}

SUs communicate opportunistically over multiple narrowband channels of throughput $W$\,kbps, that are randomly occupied by PUs (all variables are summarized in Table~\ref{tab:variables}). We consider an OSA network composed of two users where one is always the transmitter and the other one is always the receiver. The reduction of the OSA network to only two nodes allows us to neglect the impact of collisions of random access control packets on the performance of spectrum sensing algorithms. Note that the same assumption has been used in an earlier study of multi-stage sensing~\cite{jeon_twc_2008}. Our model can be extended to the multiple nodes case, however, for the brevity of the discussion, this is not addressed here. Furthermore, PUs and SUs, are assumed to arrive and depart at discrete times that are multiples of $T$, where $T$ is the slot length in seconds. We assume that the transmitting SU transmits the sensing results to the receiving SU through the dedicated control channel. Hence, the SU pair is perfectly synchronized. These two assumptions, vital in simplifying the analysis, are common in spectrum sensing literature, refer to~\cite{Park_arxiv_2009} for a good overview of related work in this area. Finally, we assume that the time taken by the SUs to switch from communicating mode to sensing mode, or from one channel to another, is negligible.

In this work we focus on OSA link throughput and PU/SU collision analysis. We assume that the obtained OSA network throughput is proportional to the duration of time the SUs use the channel and the number of channels utilized. Accordingly, throughput decreases if the SU is idle or senses the channel. Besides, in cases where the SU and the PU use the same channel simultaneously, we assume that the SU frame cannot be decoded and is considered lost. Furthermore, no acknowledgments are considered for the data transmitted by SUs. Finally, the frame error rate is assumed to be zero on PU free channels.

\begin{table}
\centering
\caption{Summary of Variables Used in the Paper}
\begin{tabular}{c|l|c}
\hline
Variable & Description & Unit \\
\hline\hline
$S$, $M$, $N$ & number of: sensing stages, sensing radios, available channels & ---\\
$B$, $b$ & buffer size, buffer utilization & --- \\
$R$, $G$ & average: SU throughput, collision probability & bps, --- \\
$W$ & channel throughput & kbps\\
$j$, $J$ & SU state vector: single narrowband radio case, parallel narrowband radio case & --- \\
$\Gamma$, $\Gamma_S$ ($\Gamma_A$) & set of SU states: feasible, where the SU is sensing (active) & ---\\
$\Theta$, $\Psi$ & set of states: single narrowband radio, parallel narrowband radio & --- \\
$\Theta_1$ ($\Theta_2$) & set of states for single narrowband radio: SU sensing, PU exists (PU absent) on channel & --- \\ 
$\theta_{\mathbf{v}}$, $\psi_{\mathbf{w}}$ & state: single narrowband radio, parallel narrowband radio & ---\\
$T$, $T_{s}$, $T_{t}$ & slot length, sensing duration at stage $S$, sensing duration for quiet and pre-sensing & (m)s, $\left(\mu\right)$s, $\left(\mu\right)$s\\
$p$, $s$ & variables denoting: PU, SU & ---\\
$i$, $x$, $k$, $l$ & supporting variables & --- \\
$c$, $m$ & indices: channel, radio & --- \\
$n_a$, $n_{\overline{a}}$ & number of channels with PU: arrivals, no arrivals & ---\\
$n_d$, $n_{\overline{d}}$ & number of channels with PU: departures, no departures & ---\\
$p_{f,s}$ ($p_{m,s}$) & probability of false alarm (mis-detection) at sensing stage & ---\\
$p_{f,t}$ ($p_{m,t}$) & probability of false alarm (mis-detection) at pre-sensing and quiet modes & ---\\
$p_{x,a}$ ($p_{x,d}$) & probability of arrival (departure) for user $x$ & ---\\
$\mathbf{w_i}$, $\mathbf{v_i}$ & vector of variables at slot $i$: single narrowband radio, parallel narrowband radio & ---\\
$\mathcal{C}_x$ & supporting vector of variables & ---\\
$\mathcal{X}$, $\mathcal{Y}$ & variables for conditional probabilities: single narrowband radio, parallel narrowband radio & ---\\
$f$ & SU traffic state for single channel SUs & ---\\
$F$ & vector of SU traffic state for parallel narrowband radio & ---\\
$I$, $\Upsilon$ & vector of PU states, set of all permutations of $I$ & ---\\
$\pi$, $\Lambda$ & vector of stationary probabilities, transition probability matrix & --- \\
$F_{N_i}$ & number of frames generated at slot $i-1$ for parallel narrowband radio & --- \\
$F_{T_i}$, $F_{B_i}$ & total number of frames at slot $i$: available for transmission, buffered & --- \\
$M_{F_i}$, $M_{A_i}$ & total number of radios at slot $i$: available for transmission, active & --- \\
$\Omega_{A_i}$, $\Omega_{I_i}$, $\Omega_{Q_i}$ & set of radios at slot $i$: active, idle, quiet & --- \\
\hline
\end{tabular}
\label{tab:variables}
\end{table}

\subsection{SU and PU Traffic Model}
\label{sec:traffic_model}

We assume that the SU sends packet traffic randomly (either with a constant or a variable bit rate), where packets are divided into frames. In the idle mode, the SU has no frames to transfer, while in the active mode, the SU is either transmitting frames or sensing the channels. The probabilities of arrival and departure of a SU frame are denoted by $p_{s,a}$ and $p_{s,d}$, respectively\footnote{Note that the arrival and departure rates of the SU frames are constant and are independent of the PU state and the SU sensing mechanism performance throughout the model. This assumption is typical for system level analysis of OSA networks. Inclusion of SU traffic which is dependent on the PU state in the model will make the observation of the SU network performance difficult, e.g. it will be hard to infer whether the decrease in throughput is a result of decreasing SU traffic or the effect of multi-stage sensing algorithm. While the inclusion of the PU-dependent SU traffic in our model is an interesting topic to consider, we leave it for future studies.}. Further, we consider SU nodes with and without data buffering capabilities. In the case of unbuffered SU nodes, incoming frames are discarded when no channels are available for transmission. In a buffered node case, a buffer of size $B$ frames is present that stores frames only when the node is in a mode of operation where sensing occupies the whole time slot. New frames are discarded if the buffer is full. The presence of the buffer increases the average throughput as incoming frames are not discarded when SUs are sensing the channel for the whole slot length.

PU channel occupancy is also time varying. That is, the probabilities of the start and end of a channel occupancy are denoted by $p_{p,a}$ and $p_{p,d}$, respectively. Finally, the average PU occupancy duration and idle times are assumed to be the same for every PU channel.

\subsection{Multi-Stage Sensing Algorithms}
\label{sec:sensing_stages}

When a SU transmitter generates a new packet, it first sends a request to send the packet on the dedicated control channel to the SU receiver. Both the sender and the receiver are constantly tuned to the control channel during the idle mode. The connection establishment time is assumed to be small in comparison to the subsequent transmission time and can therefore be neglected. 

Once connection is established, the sending SU node starts with the first stage of the multi-stage sensing algorithm on the first PU channel from a list of available PU channels. In every stage, the SU transmitter first senses the channel for $T_{s} < T$ seconds (except in the special cases, explained later in this section, where sensing occupies the whole time slot). Immediately after sensing, in the same slot, a SU frame of duration $T- T_{s}$ seconds is sent regardless of the sensing procedure outcome\footnote{Note that contrary to~\cite{jeon_twc_2008,jeon_tvt_2010} we do not assume a varying sensing time at each stage. Varying sensing time would complicate the packet forwarding process as frames always have a fixed length and cannot adapt to varying sensing duration.}. If the outcome of the sensing procedure at stage $j$ indicates that a PU is possibly present, the SU nodes proceed to the next sensing stage. If the outcome of the sensing procedure at stage $j$ indicates that the PU is possibly absent, the SUs start a new multi-stage sensing cycle.

The probabilities of false alarm and mis-detection for any of the $S$ stages of multi-stage sensing are denoted by $p_{f,s}$ and $p_{m,s}$, respectively. Note that, generally, increasing $T_s$ decreases $p_{f,s}$ and $p_{m,s}$. On the other hand, this leads to a decrease in the SUs throughput, as more time is spent on sensing for PU activity. Also note that SUs can transmit during each stage of the sensing cycle even though the PU is present as a result of a mis-detection, which is also the case for single-stage sensing algorithms.

The multi-stage spectrum sensing algorithm might outperform the single-stage sensing counterpart as a result of increased false alarm probability. In single-stage sensing algorithms, a single false alarm forces the SU to stop transmission immediately, while in the case of multi-stage spectrum sensing, the SU checks in $S$ consecutive sensing stages if a PU is actually on the channel without stopping the on-going communication. This might improve the SUs' utilization of channel vacancies specially in scenarios where the false alarm probability is high and/or the PU traffic is slow. This is further explored in Section~\ref{sec:numerical_results}.

In the multi-stage spectrum sensing algorithm, the SU transmission continues on the respective channel until the SU detects the PU presence in all $S$ stages of sensing till the last stage. The SUs may then switch to the next channel and restart the multi-stage spectrum sensing algorithm, or stay on the same channel in a prolonged sensing stage that involves no transmission, depending on the radio architecture as explained in the following sections.

\subsubsection{Multi-channel Sensing using a Single Narrowband Radio}
\label{sec:narrowband_case}

First, we focus on OSA nodes with only one radio for sensing or communication which can operate only on one frequency band at a time. SUs can switch to a different narrow frequency band when they decide to vacate their current band due to the possible presence of a PU.

The proposed model serves as a general framework enabling the analysis of any multi-stage sensing algorithm. Due to the infinite number of possibilities of multi-stage sensing algorithms, we focus our analysis on the algorithm presented in the IEEE 802.22 standard~\cite{ieee80222}, as well as our proposed extensions. The basic algorithm operates as follows. When the SUs reach the last stage of multi-stage sensing and the outcome of the sensing procedure indicates that a PU is possibly present, the SUs switch to a prolonged channel observation stage. We denote this stage by the \emph{quiet mode}. In the quiet mode, the SUs stay on that channel for the whole slot duration $T_{t} = T$, to sense for PUs. If the outcome of the sensing procedure at the quiet mode indicates that a PU is possibly present, the SUs switch to another PU channel and begin a new multi-stage sensing cycle. Otherwise, the SUs stay on the same channel and restart the multi-stage sensing cycle to detect incoming PUs or a PU that was mis-detected. Regarding the quiet mode, a false alarm with probability $p_{f,t}$, or a mis-detection with probability $p_{m,t}$ can happen (just as in~\cite{jeon_tvt_2010}, but contrary to~\cite{jeon_twc_2008}, where detection at this mode of operation is assumed perfect). Since $T_t > T_s$, then $p_{f,t}$ and $p_{m,t}$ can be chosen to be less than $p_{f,s}$ and $p_{m,s}$, respectively. This means that the sensing procedure for the prolonged channel observation stage, also called fine sensing in the IEEE 802.22 standard~\cite{ieee80222}, has higher precision than that for stages $1$ to $S$. However, this comes at the expense of a potential decrease in channel utilization by the SU. The motivation behind the quiet mode is to decrease the probability of the SU leaving a vacant channel based on false alarms (the impact of the quiet mode on the multi-stage spectrum sensing performance is discussed in more detail in Section~\ref{sec:numerical_results}).

In addition to the aforementioned algorithm, we propose and analyze algorithms which involve the usage of a prolonged channel observation period at the beginning of the sensing cycle. We denote this prolonged sensing period by the \emph{pre-sensing mode}. The length of this period is $T_{t} = T$, exactly as in the quiet mode. The motivation behind the pre-sensing mode is to decrease the collision probability with the PU, since $p_{m,t}$ is small, when the SUs switch to a new channel. At the end of pre-sensing mode, the SUs enter the first stage of sensing (if no PU has been detected or a PU has been mis-detected) or switches to a new channel and starts a new sensing procedure with pre-sensing (if a PU has been detected or a false alarm occurred). The quiet and pre-sensing stages have a considerable effect on throughput and collisions as our results demonstrate. Accordingly, we introduce and analyze algorithms combining all permutations of these modes of operation.

\subsubsection{Multi-channel Sensing using Parallel Narrowband Radios}
\label{sec:Parallel_Channel_Communication_Case_System_model}

Apart from the single narrowband radio case, we consider a system, where both SUs have $M$ narrowband radios that can operate in parallel independently. That is, at any given time, each narrowband radio can be idle, sensing or communicating regardless of the state of the other radios. Besides, the radios may operate on non-contiguous frequency bands. In this model we assume that the number of available radios is equal to the number of available channels, $N$, which is the only relevant case in practice. The incoming frames are divided across the radios with no prioritization. Each radio follows the basic single channel multi-stage sensing algorithm described in the earlier section, except once the quiet mode is reached, the radios stay in the quiet stage until the outcome of the sensing procedure indicates that the channel is vacant. If, in the quiet mode, the channel is correctly detected to be vacant or a PU occupying the channel is mis-detected, the sensing procedure outcome would indicate that the channel is vacant and the corresponding radio proceeds to the first stage of the spectrum sensing algorithm.

\section{Analytical Model}
\label{sec:analysis}

We present the foundation for the analytical model for the single and parallel narrowband radio cases separately. We divide the discussion into the Markov chain state definitions, state transition probabilities, stationary probability calculation and metrics calculation.

\subsection{Multi-channel Sensing using a Single Narrowband Radio}
\label{sec:narrowband_su}

\subsubsection{State and State Transition Definitions}
\label{sec:states_narrowband}

Let $\theta_{\mathbf{v}}$ denote the state of the Markov chain, where $\mathbf{v}=\{I,f,j,b,c\}$. In vector $\mathbf{v}$, $c \in\{1,\cdots,N\}$ is the operating channel index, which denotes the channel the SU is operating on, or was operating on before proceeding to the idle mode. SU traffic status is represented by $f$, where $f = 1$ when the SU has a new frame to send, and $f=0$ otherwise. The SU mode of operation is represented by $j \in \Gamma$, where $\Gamma$ is defined differently for each single narrowband radio sensing algorithm. For the proposed algorithms, $j = 0$ implies that the SU is idle and $j \in \{1,\cdots,S\}$ indicates that the SU is in sensing stage $j$ of multi-stage sensing, while $j = S + 1$ and $j = S + 2$ are the examples of the special cases that in the context of this work indicate that the SU is in the quiet mode and the pre-sensing mode, respectively. Define $\Gamma_S$ as a subset of $\Gamma$ representing the modes of operation where the SU is performing multi-stage sensing. Define $\Gamma_A$ as a subset of $\Gamma$ representing the modes of operation where the SU is active, that is, the SU is either performing multi-stage sensing or is in the pre-sensing or quiet mode. $\Gamma_S = \{1,\cdots,S\}$ for all algorithms, however, $\Gamma_A$ depends on the specific algorithm. The buffer level at the SU is indicated by $b \in \{0,\cdots,B\}$. Finally, the PU status are described by $I$, $|I|=N$, where $I(x)=1$ when a PU operates on channel $x$, and $I(x)=0$ otherwise. When transitioning from time slot $i$ to time slot $i+1$, the state of the Markov chain at slot $i+1$ (the SU mode of operation, buffer level and operating channel in specific) depends on the outcome of the sensing procedure that tests the existence of a PU during time slot $i$ (not time slot $i+1$). Hence, for mathematical convenience, $I$ is defined to describe the PU status in the previous time slot and not the current time slot.

The transition probability from state $\theta_{\mathbf{v_1}}$ to state $\theta_{\mathbf{v_2}}$, where $\mathbf{v_x}=\{I_x,f_x,j_x,b_x,c_x\}$ and subscript $x=\{1,2\}$ for $c$, $I$, $f$, $j$, $b$, denotes current and next time slots, respectively, is expressed as
\begin{equation}
{\Pr}\left(\mathcal{X}\right) = U\left(X\right)\displaystyle\prod_{k=1}^3{\Pr}_{k}\left(\mathcal{X}\right),
\label{eq;nb_1}
\end{equation}
where again, for notational convenience, $\mathcal{X}\triangleq\left\{\theta_{\mathbf{v_2}} \mid \theta_{\mathbf{v_1}}\right\}$. In (\ref{eq;nb_1}), ${\Pr}_1(\cdot)$ denotes the probability of PU status change, ${\Pr}_2(\cdot)$ denotes the probability of SU traffic status change, ${\Pr}_3(\cdot)$ denotes the probabilities of the sensing procedures outcomes, and $U(x)$ is the transition feasibility function and equals to one if $x$ belongs to a specified set of feasible state transitions and $U(x)=0$ otherwise. Since the traffic of the PU and SU is independent of the sensing algorithm, $\Pr_1\left(\cdot\right)$ and $\Pr_2\left(\cdot\right)$ are the same for all parallel narrowband radio algorithms, described later in Section~\ref{sec:wideband_case}. However, as $\Pr_3\left(\cdot\right)$ and $U\left(\cdot\right)$ describe transitions in the radios mode of operation, they are unique to the sensing algorithm and are described in Section~\ref{sec:single_radio_example}. The transition probabilities are consequently used in calculating the stationary probabilities as described in the following section.

\subsubsection{Stationary Probability Calculation}
\label{sec:steady_state_calculations_narrowband}

Define $\Theta$ as the set of all feasible Markov chain states. Then, define $\Lambda$ as the transition probability matrix where $\Lambda\left(k,l\right)$ denotes the transition probability from state $\Theta\left(k\right)$ to state $\Theta\left(l\right)$, where $k,l\in\{1,\cdots,\left|\Theta\right|\}$. $\Theta$ is evaluated by creating a vector of unique states, $\theta_{\mathbf{v}}$, for all feasible combinations of $c, j,b,I,f$ and directly mapping them to the values of $\Theta$. Further, define $\pi$ as the vector of stationary probabilities where $\pi\left(k\right)$ is the stationary probability of state $\Theta\left(k\right)$. Then $\pi$ can be evaluated by solving $\pi = \pi \Lambda$, knowing that $\sum_{i=1}^{\left|\Theta\right|} \pi\left(i\right) = 1$. The stationary probabilities are then used to calculate the desired metrics.

\subsubsection{Performance Metrics Calculation}
\label{sec:performance_metrics_calculations_narrowband}

Based on the stationary probability vector, $\pi$, we compute the average throughput obtained by the secondary network, $R$, and the expected number of collisions between the transmitting SU and PUs per time slot, $G$. Define $\Theta_1$ as the set of states $\theta_{\mathbf{v}}$, with $j \in \Gamma_S$ and $I\left(c\right)=1$, i.e. $\Theta_1$ is the set of states where the SU is performing multi-stage sensing and a PU was present on the channel on which the SU was operating during the previous time slot. Define $\Theta_2$ as the set of states $\theta_{\mathbf{v}}$, with $j \in \Gamma_S$ and $I\left(c\right)=0$. In calculating $R$, we assume that the SU frame cannot be decoded in case of a collision between the SU and a PU, and accordingly, $R$ can be expressed as
\begin{equation}
R = W{T-T_{s}\over{T}} \left(\sum_{\Theta\left(i\right) \in \Theta_1} {\pi\left(i\right) p_{p,d}} + \sum_{\Theta\left(i\right) \in \Theta_2} {\pi\left(i\right) \left(1-p_{p,a}\right)}\right),
\label{eq;nb_met1}
\end{equation}
where $(T-T_{s})/T$ is the portion of a time slot SU spent in transmission. Besides, $G$ can be expressed as
\begin{equation}
G = \sum_{\Theta\left(i\right) \in \Theta_1} {\pi\left(i\right) \left(1-p_{p,d}\right)} + \sum_{\Theta\left(i\right) \in \Theta_2} {\pi\left(i\right) p_{p,a}}.
\label{eq;nb_met2}
\end{equation}
Note that in (\ref{eq;nb_met1}) and (\ref{eq;nb_met2}), the information regarding the PU status according to $\theta_{\mathbf{v}}$, given by $I$, pertains to that of the previous time slot. Accordingly, the probability that no PUs exist at the current time slot at channel $c$ given that $I(c) = 1$ equals $p_{p,d}$, i.e. the probability that the PU user that was utilizing channel $c$ in the previous time slot is not active during the current time slot. Likewise, the probability that a PU is active on channel $c$ given that $I(c) = 1$ is $(1-p_{p,d})$. This accounts for the $p_{p,d}$ and $(1-p_{p,a})$ terms in (\ref{eq;nb_met1}) and the $p_{p,a}$ and $(1-p_{p,d})$ terms in (\ref{eq;nb_met2}). In the following sections we will describe probabilities governing (\ref{eq;nb_1}) for each sensing algorithm.

\subsubsection{Definitions of ${\Pr}_{1}(\cdot)$ and ${\Pr}_{2}(\cdot)$}
\label{sec:definition_pr1_pr2_narrowband}

Let $n_a$ denote the number of channels with PU arrivals, $n_{\overline{a}}$ denote the number of channels with no PU arrivals, $n_d$ denote the number of channels with PU departures and $n_{\overline{d}}$ denote the number of channels that remained occupied by the PU. It follows that $n_{a} = \overline{I}_1\circ I_2$, $n_{\overline{a}} = \overline{I}_1\circ \overline{I}_2$, $n_d = I_1\circ \overline{I}_2$, and $n_{\overline{d}} = I_1\circ I_2$, where $\overline{x}$ is the bitwise binary complement of $x$ and "$\circ$ is the base-10 dot product operator. Accordingly, ${\Pr}_1(\cdot)$ is expressed as
\begin{equation}
{\Pr}_1\left(\mathcal{X}\right)=p_{p,a}^{n_a}(1-p_{p,a})^{n_{\overline{a}}}p_{p,d}^{n_d}(1-p_{p,d})^{n_{\overline{d}}},
\label{eq;nb_2}
\end{equation}
where $I_1, I_2 \in \Upsilon$ and $\Upsilon$ is the set of all $2^N$ permutations of $I$. ${\Pr}_2(\cdot)$ is expressed as
\begin{equation}
{\Pr}_2\left(\mathcal{X}\right) = 
\begin{cases}
1-p_{s,a}, & f_1 = f_2 = 0,\\
p_{s,a}, & f_1 = 0, f_2 = 1,\\
p_{s,d}, & f_1 = 1, f_2 = 0,\\
1-p_{s,d}, & f_1 = f_2 = 1.
\end{cases}
\label{eq;nb_3}
\end{equation}

\subsection{Multi-channel Sensing using Parallel Narrowband Radios}
\label{sec:wideband_case}

\subsubsection{State and State Transition Definitions}
\label{sec:states_wideband}

Let $\psi_{\mathbf{w}}$ denote the states of the Markov chain, where $\mathbf{w}=\{I,F,J,b\}$. In $\mathbf{w}$, $I$ and $b$ are defined the same as for the single narrowband radio case. Let $\psi_{\mathbf{w_i}}$ denote the Markov chain state at time slot $i$, where $\mathbf{w_i}=\{I_i,F_i,J_i,b_i\}$. In the context of our work, the parallel channel SU is assumed to have $M$ single channel radios, the radio states are described by $J$, $|J|=M$, where $J\left(x\right) \in \Gamma$ describes the mode of operation of radio $x$. For the considered algorithm $\Gamma = \{0,\cdots,S+1\}$ and $J\left(x\right) = S+1$ indicates that radio $x$ is in the quiet mode. The rest of the radio modes of operation are defined as for the single narrowband radio case. Furthermore, $\Gamma_S = \{1,\cdots,S\}$, as in the case of singe parallel radios, defines the subset of $\Gamma$ that represents the modes of operation where the SU radio is performing multi-stage sensing.

SU traffic is assumed to be in the form of a frame stream with frame arrival and departure probabilities as in the single narrowband radio case. The difference is that up to $M$ frames can be generated per time slot. Thus, for analysis, a slot can be hypothetically segmented to $M$ parts where a new frame can be generated at each part. The state of these frames is represented by $F_i$, $|F_i|=M$, where $F_i\left(x\right) = 1$ if a frame was created at division $x$ of slot $i-1$ and $F_i\left(x\right) = 0$ otherwise. Hence, the total number of frames generated at slot $i-1$, denoted by $F_{N_i}$, can be expressed as $F_{N_i} = \sum_{x=1}^{M}{F_i\left(x\right)}$. These frames are accumulated with buffered frames from time slot $i-1$ and are ready for transmission at time slot $i$. The excess frames that cannot be transmitted at time slot $i$ due to the limited number of available channels are buffered where a maximum of $B$ frames can be stored. The total number of frames available for transmission at time slot $i$, denoted by $F_{T_i}$, can be expressed as $F_{T_i} = b_{i-1} + F_{N_i}$ where $b_{i-1}$ is the number of buffered frames from time slot $i-1$. Radios that are in the quiet mode at slot $i$ are not used for frame transmission. Hence, the total number of radios available for frame transmission at slot $i$ can be expressed as $M_{F_i} = \sum_{m=1}^{M}{V\left(J_i\left(m\right)\right)}$ where $V\left(J_i\left(m\right)\right) = 1$ when $J_i\left(m\right)\neq S+1$ and $V\left(J_i\left(m\right)\right)=0$, otherwise. Accordingly, the total number of radios that will be active at slot $i$, which equals the total number of frames that will be transmitted, is $M_{A_i} = \min\{F_{T_i},M_{F_i}\}$. These radios are selected from the set of radios that are available for frame transmission at slot $i$ in ascending order of the radio indices (due to the symmetry of the system). 

Denote the set of radios that will be active at slot $i$ by $\Omega_{A_i}$ where $|\Omega_{A_i}| = M_{A_i}$. If $M_{F_i} > M_{A_i}$, the extra radios that are available for frame transmission at slot $i$ will be idle. Denote the set of these radios by $\Omega_{I_i}$ where $|\Omega_{I_i}| = \max\{0, M_{F_i}-F_{T_i}\}$. Radios that are in the quiet mode will stay in the quiet mode until the outcome of the sensing procedure indicates the absence of a PU. Denote the set of radios that are in the quiet mode at slot $i$ by $\Omega_{Q_i}$ where $|\Omega_{Q_i}| = M-M_{F_i}$. If $M_{F_i} \geq F_{T_i}$, then $M_{A_i} = F_{T_i}$ and all available frames will be transmitted, thus, $b_i = 0$. Otherwise, $M_{A_i} < F_{T_i}$ and only $M_{A_i}$ frames will be transmitted, hence, $b_i = \min\{B, F_{T_i}-M_{A_i}\}$. The rest of the frames, if any, will be dropped. The transition probability from state $\psi_{\mathbf{w_1}}$ to state $\psi_{\mathbf{w_2}}$ is given by
\begin{equation}
{\Pr}\left(\mathcal{Y}\right) = U\left(\mathcal{Y}\right)\displaystyle\prod_{k=1}^3 {\Pr}_k{\left(\mathcal{Y}\right)},
\label{eq;wb_4}
\end{equation}
where, for notational convenience, $\mathcal{Y}\triangleq\left\{\psi_{\mathbf{w_2}} | \psi_{\mathbf{w_1}}\right\}$. In (\ref{eq;wb_4}), $\Pr_1\left(\cdot\right)$ represents the probability of PU status change, $\Pr_2\left(\cdot\right)$ denotes the probability of SU traffic status change, $\Pr_3\left(\cdot\right)$ represents the probabilities of the sensing procedures outcomes, and $U\left(\cdot\right)$ is the transition feasibility function and $U(x)=1$ if $x$ belongs to a specified set of feasible state transitions and $U(x)=0$ otherwise. Since the traffic of the PU and SU is independent of the sensing algorithm, $\Pr_1\left(\cdot\right)$ and $\Pr_2\left(\cdot\right)$ are the same for all algorithms. However, as $\Pr_3\left(\cdot\right)$ and $U\left(\cdot\right)$ describe transitions in the radios mode of operation, they are unique to the sensing algorithm and are described in Section~\ref{sec:parallel_radio_example}.

\subsubsection{Stationary Probability Calculation}
\label{sec:steady_state_calculations_wideband}

Denote $\Psi$ as the set of all feasible states. $\Psi$ is evaluated by creating a vector of unique states, $\psi_{\mathbf{w}}$ for all feasible combinations of $J,b,I,F$ and directly mapping them to the values of $\Psi$. Define $\pi$ as the vector of stationary probabilities where $\pi\left(k\right)$ is the stationary probability of state $\Psi\left(k\right)$, where $k \in\{1,\cdots,\left|\Psi\right|\}$. Then $\pi$ can be evaluated by solving $\pi = \pi \Psi$, knowing that $\sum_{i=1}^{\left|\Psi\right|} \pi\left(i\right) = 1$.

\subsubsection{Performance Metrics Calculation}
\label{sec:performance_metrics_calculations_wideband}

The throughput, $R$, can be expressed as
\begin{align}
R = W{T-T_{s}\over{T}} \sum_{k=1}^{\left|\Psi\right|} { \sum_{m=1}^M{ \pi\left(k\right) \left[ p_{p,d} U_a(J_k\left(m\right), I_k\left(m\right)) + \left(1-p_{p,a}\right)U_b(J_k\left(m\right), I_k\left(m\right))\right]}},
\label{eq;wb_met1}
\end{align}
where $U_a\left(j,i\right) = 1$ if $j \in \Gamma_S \text{ and } i = 1$, and zero otherwise; $U_b\left(j,i\right) = 1$ if $j \in \Gamma_S \text{ and } i = 0$, and zero otherwise. $J_k\left(m\right)$ and $I_k\left(m\right)$ are $J$ and $I$ for radio $m$ at state $k$, respectively. 

The expected number of collisions, $G$, can be expressed as
\begin{align}
G = \sum_{k=1}^{\left|\Psi\right|} { \sum_{m=1}^M{ \pi\left(k\right) \left[ \left(1-p_{p,d}\right) U_a(J_k\left(m\right), I_k\left(m\right)) + p_{p,a}U_b(J_k\left(m\right), I_k\left(m\right))\right]}}.
\label{eq;wb_met2}
\end{align}

\subsubsection{Definitions of ${\Pr}_{1}(\cdot)$ and ${\Pr}_{2}(\cdot)$}
\label{sec:definition_pr1_pr2_narrowband}

$\Pr_1\left(\cdot\right)$ is defined as in (\ref{eq;nb_2}). The expression for ${\Pr}_2(\cdot)$ is an extension to that of the single channel case. Hence,
\begin{equation}
{\Pr}_2\left(\mathcal{Y}\right) = {\Pr}_S\left(F_2\left(1\right) | F_1\left(M\right)\right)\displaystyle\prod_{m=1}^{M-1}{{\Pr}_S\left(F_2\left(m+1\right)| F_2\left(m\right)\right)},
\label{eq;wb_5}
\end{equation}
where ${\Pr}_S(\cdot)$ is defined as (\ref{eq;nb_3}) replacing $f_1$ with $x$ and $f_2$ with $y$.

\section{Multi-Stage Spectrum Sensing Algorithm Examples: Analysis}
\label{sec:design_examples}

We present different examples of multi-stages sensing algorithms. As mentioned in Section~\ref{sec:states_narrowband}, ${\Pr}_3(\cdot)$ and $U(\cdot)$ are unique to the sensing algorithms and are thus derived here.

\subsection{Multi-channel Sensing using a Single Narrowband Radio}
\label{sec:single_radio_example}

Before proceeding with the description of ${\Pr}_3(\cdot)$ and $U(\cdot)$, we denote transitions where the SU stays on the same channel as $\mathcal{C}_1\triangleq\{c_1 = c_2\}$. Further, denote transitions where the SU switches to the next channel as $\mathcal{C}_2\triangleq\{c_1 < M, c_2=c_1+1 \text{ or } c_1= M, c_2=1\}$.

\vspace{-0.1cm}

\paragraph{No Pre-sensing, Quiet mode (P$_0$Q$_1$)}

This is the algorithm presented in the IEEE 802.22 standard~\cite{ieee80222}. Since it features a quiet mode but no pre-sensing mode, thus $\Gamma = \{0,\cdots,S+1\}$ and $\Gamma_A = \{1,\cdots,S+1\}$ (note that $\Gamma_S=\{1,\ldots,S\}$ is defined in the same way as in Section~\ref{sec:states_narrowband}). Then
\begin{subnumcases}
{\label{nb_p3}
{\Pr}_3\left(\mathcal{Y}\right) = }
1, & $j_1=0 \text{ or } j_2 = 0,\mathcal{C}_1,$\label{nb_p3a} \\
p_{f,s}, 	 & $I_2(c_1) = 0, j_1\in \Gamma_S, j_2=j_1+1,\mathcal{C}_1,$\label{nb_p3b} \\
p_{f,t}, 	 & $I_2(c_1) = 0, j_1 = S + 1, j_2 = 1, \mathcal{C}_2,$\label{nb_p3c} \\
1 - p_{f,s}, & $I_2(c_1) = 0,j_1\in \Gamma_S, j_2=1,\mathcal{C}_1,$\label{nb_p3d} \\
1 - p_{f,t}, & $I_2(c_1) = 0,j_1 = S + 1, j_2=1,\mathcal{C}_1,$\label{nb_p3e} \\
1 - p_{m,s}, & $I_2(c_1) = 1,j_1\in \Gamma_S, j_2=j_1+1,\mathcal{C}_1,$\label{nb_p3f} \\
1 - p_{m,t}, & $I_2(c_1) = 1, j_1 = S + 1, j_2 = 1, \mathcal{C}_2,$\label{nb_p3g} \\
p_{m,s}, 	 & $I_2(c_1) = 1,j_1\in \Gamma_S, j_2=1, \mathcal{C}_1,$\label{nb_p3h} \\
p_{m,t}, 	 & $I_2(c_1) = 1,j_1 = S + 1, j_2=1, \mathcal{C}_1,$\label{nb_p3i} \\
0, & $\text{otherwise}$.
\end{subnumcases}
We briefly explain the conditions (\ref{nb_p3a})--(\ref{nb_p3i}). Transitions to and from the idle mode do not involve sensing, hence, ${\Pr}_3(\cdot) = 1$ for transitions where $j_1=0$ or $j_2 = 0$ as shown in (\ref{nb_p3a}). The cases where the SU is at any sensing stage and proceeds to the next stage, based on a false alarm or a successful detection of a PU, are represented by (\ref{nb_p3b}) and (\ref{nb_p3f}), respectively. The cases where the SU is in the quiet mode and switches to another channel, based on a false alarm or a successful detection of a PU, are represented by (\ref{nb_p3c}) and (\ref{nb_p3g}), respectively. The cases where the SU proceeds from an active mode of operation to the first stage of sensing, based on correctly detecting the absence of PUs, are represented by (\ref{nb_p3d}) and (\ref{nb_p3e}). The cases where the SU proceeds from an active stage to the first stage of sensing, based on a mis-detection of a PU, are represented by (\ref{nb_p3h}) and (\ref{nb_p3i}), respectively. 

Denote transitions where the buffer level remains unchanged and a new frame is produced at the next state as $\mathcal{C}_3\triangleq\{b_1 = b_2, f_2 = 1\}$. Denote transitions where the buffer level decreases by one frame and no new frames are produced at the next state as $\mathcal{C}_4\triangleq\{b_2 = b_1 - 1, b_1 > 0, f_2 = 0\}$. Then, feasible cases where the SU stays on the same channel, $\mathcal{C}_1$, are
\begin{subnumcases} {\label{nb1_U} U(\cdot)=1\iff}
j_1 \in \{0,\cdots,S\}, j_2 = 0, b_2=f_2=0, \label{nb1_Ua} \\
j_1 = S+1, j_2 = f_2 = B=0, \label{nb1_Ub} \\
j_1 = 0, j_2 = 1, b_2=0, f_2=1, \label{nb1_Uc} \\
j_1 \in \{1,\cdots,S-1\},j_2=j_1+1, \mathcal{C}_3 \text{ or } \mathcal{C}_4, \label{nb1_Ud} \\
j_1 = S, j_2 = S+1, f_2=1, b_2= \min(b_1+1, B), \label{nb1_Ue} \\
j_1 = S, j_2 = S+1, f_2=0, b_2= b_1, b_1>0, \label{nb1_Uf} \\
j_1 \in \Gamma_A, j_2 = 1, \mathcal{C}_3 \text{ or } \mathcal{C}_4, \label{nb1_Ug}
\end{subnumcases}
The feasible case where the SU switches to the next channel, $\mathcal{C}_2$, is
\begin{equation}
U(\cdot)=1\iff j_1 = S + 1, j_2 = 1, \mathcal{C}_3 \text{ or } \mathcal{C}_4.
\label{eq;nb_428}
\end{equation}
We explain the conditions behind $U(\cdot)$. A SU that is in the idle mode or is performing multi-stage sensing stays at or proceeds to the idle mode, respectively, if no new frames are generated and the buffer is empty, as represented in (\ref{nb1_Ua}). A SU with a buffer that is operating in the quiet mode must have at least one frame in its buffer, as otherwise, the SU should be in the idle mode. However, if the SU has no buffer, it proceeds to the idle mode if there are no new generated frames, as shown in (\ref{nb1_Ub}). An idle SU proceeds to the first stage of sensing whenever a new frame is generated, the SU transmits the frame and the buffer remains empty as represented in (\ref{nb1_Uc}). Condition (\ref{nb1_Ud}) accounts for the case when a SU is in one of the first $S-1$ sensing stages with a new generated frame and/or has at least one frame in its buffer. Then the SU may proceed to the next sensing stage (based on the outcome of the sensing procedure). Furthermore, if this SU has a new generated frame, the buffer level remains unchanged; otherwise, the buffer level decreases by one. If the SU switches to the quiet mode and has a new generated frame as shown in (\ref{nb1_Ue}), the buffer level increases by one if the buffer is not full, otherwise, the new frame is discarded. If the SU is at the last stage of sensing with at least one frame in its buffer and no new frames are generated, the SU may switch to the quiet mode (based on the sensing procedure outcome) and the buffer level stays the same as presented in (\ref{nb1_Uf}). An active SU with frames to transmit that mis-detects a PU or correctly detects the absence of a PU proceeds to the first stage of sensing. In this case, as described in (\ref{nb1_Ug}), if the SU has a new generated frame, the buffer level remains unchanged; otherwise, the buffer level decreases by one. If a SU is in the quiet mode and has frames to transmit and the sensing procedure results in a false alarm or a successful detection of a PU, the SU switches to a new channel and starts at the first stage of sensing. If this SU has a new generated frame, the buffer level remains unchanged; otherwise, the buffer level decreases by one. This transition is shown in (\ref{eq;nb_428}).

\vspace{-0.1cm}

\paragraph{No Pre-sensing, No Quiet Mode (P$_0$Q$_0$)}

This algorithm aims at reducing the time spent in sensing by omitting the pre-sensing and quiet modes of operation. When the outcome of the sensing procedure at stage $S$ indicates that a PU is possibly present, the SU switches to a new channel and restart the sensing procedure. Then, $\Gamma = \{0,1,\cdots,S\}$ and $\Gamma_A = \Gamma_S = \{1,\cdots,S\}$. Since there are no pre-sensing or quiet modes, frames do not get buffered. Accordingly, we assume the absence of a buffer, thus, $B = b_1 = b_2 = 0$. Hence
\begin{equation}
{\Pr}_3\left(\mathcal{Y}\right) = 
\begin{cases}
1, 			 \!\!\!\!& \begin{split} j_1=0 \text{ or } j_2 = 0,\mathcal{C}_1;\end{split}\\
p_{f,s}, 	 \!\!\!\!& \begin{split} &I_2(c_1) = 0, j_1 \! \in \! \{1,\cdots,S\!-\!1\}, j_2\!=\! j_1\!+\!1,\mathcal{C}_1\text{ or } \\&I_2(c_1) = 0, j_1 = S, j_2 = 1, \mathcal{C}_2;\end{split} \\
1 - p_{f,s}, \!\!\!\!& \begin{split}I_2(c_1) = 0,j_1\in \Gamma_S, j_2=1,\mathcal{C}_1;\end{split} \\
1 - p_{m,s}, \!\!\!\!& \begin{split} &I_2(c_1) = 1,j_1 \! \in \! \{1,\cdots,S\!-\!1\}, j_2\!=\! j_1\!+\!1,\mathcal{C}_1\text{ or } \\&I_2(c_1) = 1, j_1 = S, j_2 = 1, \mathcal{C}_2;\end{split} \\
p_{m,s}, 	 \!\!\!\!& \begin{split}I_2(c_1) = 1,j_1\in \Gamma_S, j_2=1, \mathcal{C}_1;\end{split} \\
0, 			 \!\!\!\!& \begin{split}\text{otherwise.}\end{split}
\end{cases}
\label{eq;nb2_p3}
\end{equation}
The derivation of ${\Pr}_3(\cdot)$ for P$_0$Q$_0$ follows that for P$_0$Q$_1$. The difference in expressions is caused by the absence of the quiet mode. Regarding $U(\cdot)$, feasible transitions where the SU stays on the same channel, $\mathcal{C}_1$, are
\begin{subnumcases} {\label{eq;nb2_U1} U(\cdot)=1\iff}
j_1 \in \Gamma, j_2 = 0, f_2=0, \label{eq;nb2_Ua} \\
j_1 \in \Gamma, j_2 = 1, f_2=1,\label{eq;nb2_Ub} \\
j_1 \in \{1,\cdots,S-1\}, j_2=j_1+1 ,f_2=1. \label{eq;nb2_Uc}
\end{subnumcases}
The feasible case where the SU switches to the next channel ($c_1 < M, c_2 = c_1 +1$ or $c_1 = M,c_2 = 1$) is
\begin{equation}
U(\cdot)=1\iff j_1 = S, j_2 = 1, f_2=1.
\label{eq;nb2_U2}
\end{equation}
Since a buffer is unnecessary for this algorithm, a SU with no new generated frames proceeds to the idle mode irrespective of its previous state as shown in (\ref{eq;nb2_Ua}). Condition (\ref{eq;nb2_Ub}) indicates that a SU with a new generated frame proceeds to the first stage of sensing in case of mis-detecting or correctly detecting the absence of a PU. In case of a false alarm or correctly detecting a PU, if the SU is at any of the first $S-1$ sensing stages, the SU stays on the same channel and proceeds to the next sensing stage as presented in (\ref{eq;nb2_Uc}). Otherwise, if the SU is at the last stage of sensing, it switches to the next channel starting at the first sensing stage as shown in (\ref{eq;nb2_U2}).

\vspace{-0.1cm}

\paragraph{Pre-sensing, Quiet Mode (P$_1$Q$_1$)}

As a complement to P$_0$Q$_0$, this algorithm features both a pre-sensing as well as a quiet mode of operation. If the outcome of the sensing procedure at the pre-sensing mode indicates that a PU is possibly present, due to a false alarm with probability $p_{f,t}$ or a sucessfull detection with probability $1-p_{m,t}$, the SU switch to the consecutive channel starting with the pre-sensing mode. Otherwise, the SU proceeds to the first stage of sensing at the same channel. In this algorithm, $\Gamma = \{0,\cdots,S+2\}$ and $\Gamma_A = \{1,\cdots,S+2\}$. The expression for ${\Pr}_3(\cdot)$ follows that for P$_0$Q$_1$ as described in (\ref{nb_p3}) but, due to the introduction of the pre-sensing mode, the condition on $j_1$ in (\ref{nb_p3c}), (\ref{nb_p3e}), (\ref{nb_p3g}) and (\ref{nb_p3i}) changes to $j_1\in\{S+1,S+2\}$. Also, the condition on $j_2$ in (\ref{nb_p3c}) and (\ref{nb_p3g}) changes to $j_2 = S+2$. The expression for $U\left(\cdot\right)$ for cases where the SUs stay on the same channel follows that for P$_0$Q$_1$ as given in (\ref{nb1_U}) with the following alterations: (\ref{nb1_Ub}) changes to $j_1 \in \{S+1,S+2\}, j_2 = f_2 = B=0$, and (\ref{nb1_Uc}) changes to $j_1 = 0, j_2 = S+2, b_2=\min(b_1+1, B), f_2=1$. For transitions where the SU switches to the next channel
\begin{equation}
U(\cdot)=1\iff
\begin{cases}
j_1 \in \{S+1,S+2\}, j_2 = S+2, f_2=1, b_2= \min(b_1+1, B),\\
j_1 \in \{S+1,S+2\}, j_2 = S+2, f_2=0, b_2=b_1,b_1>0,
\end{cases}
\label{eq;nb_99}
\end{equation}

\vspace{-0.1cm}

\paragraph{Pre-sensing, No Quiet Mode (P$_1$Q$_0$)}

To investigate the effect of introducing the pre-sensing mode of operation only, we propose and describe an algorithm that features a pre-sensing mode with no quiet mode. Hence, $\Gamma = \{0,\cdots,S,S+2\}$ and $\Gamma_A = \{1,\cdots,S,S+2\}$. The expression for ${\Pr}_3(\cdot)$ is as given in (\ref{nb_p3}) but the condition on $j_1$ in (\ref{nb_p3b}) and (\ref{nb_p3f}) changes to $j_1 \in \{1,\cdots,S-1\}$. The condition on $j_1$ in (\ref{nb_p3e}) and (\ref{nb_p3i}) changes to $j_1 = S + 2$. Finally, the conditions on $j_1$ and $j_2$ in (\ref{nb_p3c}) and (\ref{nb_p3g}) change to $j_1 \in \{S,S+2\}$ and $j_2=S+2$. This is to accommodate for the absence of the quiet mode and the presence of the pre-sensing mode.

Regarding $U(\cdot)$, conditions for transitions for which $U\left(\cdot\right) = 1$ where the SU stays on the same channel are given as (\ref{nb1_U}) with the following alterations. Condition (\ref{nb1_Ub}) is changed to $j_1 = S+2, j_2 = f_2 = B=0$. Condition (\ref{nb1_Uc}) is altered to $j_1 =0, j_2 = S+2, b_2=\min(b_1+1,B), f_2=1$. Conditions (\ref{nb1_Ue}) and (\ref{nb1_Uf}) are dropped as there is no quiet mode. Condition (\ref{nb1_Uc}) is altered to accommodate for the pre-sensing mode; an idle SU with a new generated frame switches to the pre-sensing mode and buffers the new frame if a buffer is available. Feasible transitions where the SU switches to the next channel are given as (\ref{eq;nb_99}) replacing $j_1\in\{S+1,S+2\}$ with $j_1\in\{S,S+2\}$. It implies that if the SU is at the last sensing stage or in the pre-sensing mode the SU may switch to the next channel starting in the pre-sensing mode depending on the sensing procedure outcome. Since the SU is not allowed to communicate on the channel while in the pre-sensing mode, if a new frame is generated, the frame is either buffered, if the buffer is not full, or dropped if the buffer is full or if there is no buffer. Otherwise, if no new frames are generated, the buffer level remains unchanged.

\vspace{-0.5cm}

\subsection{Multi-channel Sensing using Parallel Narrowband Radios}
\label{sec:parallel_radio_example}

For the parallel narrowband radios case, we describe one algorithm implementation, noting that derivations for other algorithms (not presented in this paper) may follow from Section~\ref{sec:single_radio_example}. ${\Pr}_3(\cdot)$, can be expressed as
\vspace{-0.5cm}
\begin{equation}
{\Pr}_{3}\left(\mathcal{Y}\right) = \displaystyle\prod_{m_1\in \Omega_{A_1}}{{\Pr}_{3,1}\left(\mathcal{Y}\right)} \displaystyle\prod_{m_2\in \Omega_{Q_1}}{{\Pr}_{3,2}\left(\mathcal{Y}\right)} \displaystyle\prod_{m_3\in \Omega_{I_1}}{{\Pr}_{3,3}\left(\mathcal{Y}\right)},
\label{eq;wb_7}
\end{equation}
where ${\Pr}_{3,1}(\cdot)$, ${\Pr}_{3,2}(\cdot)$, and ${\Pr}_{3,3}(\cdot)$ represent the transition probabilities for radios that are elements of sets $\Omega_{A_1}$, $\Omega_{Q_1}$, and $\Omega_{I_1}$, introduced in Section~\ref{sec:states_wideband}, respectively. ${\Pr}_{3,1}(\cdot)$ can be expressed as
\begin{equation}
{\Pr}_{3,1}\left(\mathcal{Y}\right)=
\begin{cases}
1, & \begin{split}&J_1\left(m\right)\in\{1,\cdots,S-1\},J_2\left(m\right)=0;\end{split}\\
p_{f,s}, & \begin{split}&I_2\left(m\right)=0,J_1\left(m\right)\in\Gamma_S,\\&J_2\left(m\right)=J_1\left(m\right)+1;\end{split}\\
1-p_{m,s}, & \begin{split}&I_2\left(m\right)=1,J_1\left(m\right)\in\Gamma_S,\\&J_2\left(m\right)=J_1\left(m\right)+1;\end{split}\\
1-p_{f,s}, & \begin{split}&I_2\left(m\right)=0,J_1\left(m\right)\in\Gamma_S, J_2\left(m\right)=1\text{ or } \\&I_2\left(m\right)=0,J_1\left(m\right)=S, J_2\left(m\right)=0;\end{split}\\
p_{m,s}, & \begin{split}&I_2\left(m\right)=1,J_1\left(m\right)\in\Gamma_S, J_2\left(m\right)=1\text{ or } \\&I_2\left(m\right)=1,J_1\left(m\right)=S, J_2\left(m\right)=0;\end{split}\\
0, & \begin{split}\text{otherwise}.\end{split}
\end{cases}
\label{eq;wb_8}
\end{equation}
The expressions shown in (\ref{eq;wb_8}) follow that for (\ref{nb_p3}) where they express whether the active radio proceeds to the next sensing stage or back to the first stage of sensing based on the sensing decision and the presence or absence of a PU. The only difference is for cases where the radio switches to the idle mode. A radio in any of the first $S-1$ sensing stages can switch to the idle mode in case it is not needed for frame transmission. The outcome of the sensing procedure is neglected. For a radio that is in the last stage of sensing and is not needed for frame transmission, the radio can switch to the idle mode only if the sensing procedure does not detect a PU. Otherwise, the radio switches to the quiet mode. ${\Pr}_{3,2}(\cdot)$ can be expressed as
\begin{equation}
{\Pr}_{3,2}\left(\mathcal{Y}\right) = 
\begin{cases}
p_{f,t}, & I_2\left(m\right)=0,J_2\left(m\right)=S+1;\\
1-p_{m,t}, & I_2\left(m\right)=1,J_2\left(m\right)=S+1;\\
1-p_{f,t}, & \begin{split} & I_2\left(m\right)=0,J_2\left(m\right)=0 \text{ or } \\ & I_2\left(m\right)=0,J_2\left(m\right)=1;\end{split}\\
p_{m,t},   & \begin{split} & I_2\left(m\right)=1,J_2\left(m\right)=0 \text{ or } \\ & I_2\left(m\right)=1,J_2\left(m\right)=1;\end{split}\\
0, & \text{otherwise}.
\end{cases}
\label{eq;wb_9}
\end{equation}
The expressions describe the state transitions for radios in the quiet mode. In case of a false alarm or a successful detection of a PU, the radio stays in the quiet mode. Otherwise, the radio proceeds either to the idle mode, in case it is not needed for frame transmission, or to the first stage of spectrum sensing. ${\Pr}_{3,3}(\cdot)$ can be expressed as
\begin{equation}
{\Pr}_{3,3}\left(\mathcal{Y}\right) = 
\begin{cases}
1, & J_2\left(m\right)=0 \text{ or } J_2\left(m\right)=1,\\
0, & \text{otherwise}.
\end{cases}
\label{eq;wb_10}
\end{equation}
This expression is for radios that are currently in the idle mode. These radios can either stay in the idle mode, if they are not needed for transmission, or proceed to the first stage of spectrum sensing, otherwise. No spectrum sensing is done by idle radios. Conditions for feasible state transitions are given by
\begin{equation}
U(\cdot)=1\iff
F_{T_2} \geq M_{A_2}, b_2 = \min\{F_{T_2}-M_{A_2}, B\}.
\label{eq;wb_11}
\end{equation}
These conditions indicate that the number of active radios should be less than or equal to the number of frames available for transmission. Besides, since only $M_{A_2}$ frames can be transmitted, $F_{T_2}-M_{A_2}$ frames will need to be buffered. However, as the buffer can only hold $B$ frames, the total number of frames that will be buffered equals $\min\{F_{T_2}-M_{A_2}, B\}$.

\section{Numerical Results}
\label{sec:numerical_results}

Due to the vast number of parameters in the proposed model, we focus on the results that capture the interplay between PU and SU traffic and sensing algorithms. In specific, we present numerical results showing throughput and collision probability as a function of the number of sensing stages. Moreover, we investigate the relationship between the sensing time and the performance metrics. Furthermore, we demonstrate the relationship between buffer size and the obtained throughput. Finally, we compare the performance of single and parallel narrowband radio architectures. The results for all proposed algorithms are shown for single and parallel narrowband radio architectures.

We adopt a set of parameters and assumptions that are common to all scenarios presented in this section, unless otherwise stated. Specifically, we adopt energy detection as the physical layer sensing technique. Furthermore, we assume an AWGN channel, where the expressions for the probabilities of false alarm and mis-detection for such channel can be found in~\cite[Eq. (12)]{Digham} and~\cite[Eq. (14)]{Digham}, respectively\footnote{Note that our model is applicable to any channel model, including the ones discussed in~\cite{jeon_twc_2008,jeon_tvt_2010}, provided that there is an analytical expression linking the sensing time and the probabilities of false alarm and detection for a given SNR.}. The bandwidth for each PU channel is set to 6\,MHz, as in~\cite[Sec. IV]{jeon_tvt_2010},~\cite[Sec. IV]{jeon_twc_2008}, the channel throughput $W=1$\,Mbps and the SNR at the sensing receiver is set to --10\,dB. Slot time is set to $T=1$\,ms, as assumed in e.g.~\cite{Park_arxiv_2009}. Note that we verify all analytical results presented in this section using Monte Carlo simulations.

\subsection{Single Narrowband Radio: Throughput and Collisions versus Number of Stages}
\label{sec:sensing_stages_results}

\begin{table}
\caption{Summary of Common Parameters used in Numerical Evaluation}
\begin{center}
\begin{tabular}{c|c}
\hline
Descriptive Name & Parameter Values\\
\hline\hline
Long-$T_s$ Low-$p_f$ & $T_s=0.24T$, $p_{f,s}=0.1$\\
Short-$T_s$ High-$p_f$ &  $T_s=0.1T$, $p_{f,s}=0.36$\\ 
Slow PU traffic & $p_{p,a}=p_{p,d}=0.01$\\
Fast PU traffic & $p_{p,a}=0.5$, $p_{p,d}=0.1$\\
\hline
\end{tabular}
\end{center}
\label{tab:parameters}
\end{table}

We investigate the effect of varying the number of sensing stages on throughput and collision probability for each sensing algorithm in the single narrowband radio architecture case. We assume $N=6$ channels, which represents a typical value for multi-channel networks, cf. $N\in\{2,\ldots,12\}$~\cite{Park_arxiv_2009}. For the SU traffic, we assume $p_{s,a} = 1$ and $p_{s,d} = 0$, i.e the SU transmits constant bit rate traffic, which saturates the channel. We consider SUs without buffer, since our analysis has shown that buffer size does not impact throughput when SUs are always transmitting frames. We consider two cases, described below, related to sensing to explore the tradeoff between the sensing time and the resulting throughput. We achieve this by varying the sensing time and the energy detection sensing threshold while keeping the probability of mis-detection constant at $p_{m,s}=0.1$. The selected values for $p_{f,s}$ are 0.1 and 0.36 which correspond to $T_s = 0.24 T$ and $T_s = 0.1 T$, respectively. We denote the former sensing technique by long-$T_s$ low-$p_f$ and the latter by short-$T_s$ high-$p_f$. The energy detection sensing thresholds for the quiet and pre-sensing modes of operation are the same as those for the other stages of sensing. Two realistic PU traffic models are considered: (i) one with slow PU traffic (relative to the slot duration) with $p_{p,a} = p_{p,d} = 0.01$, and (ii) one with fast PU traffic with $p_{p,a} = 0.5$ and $p_{p,d} = 0.1$~\cite[Table 3]{wellens_phycom_2009}.
\begin{figure}
\centering
\subfigure[]{\includegraphics[width=0.30\columnwidth]{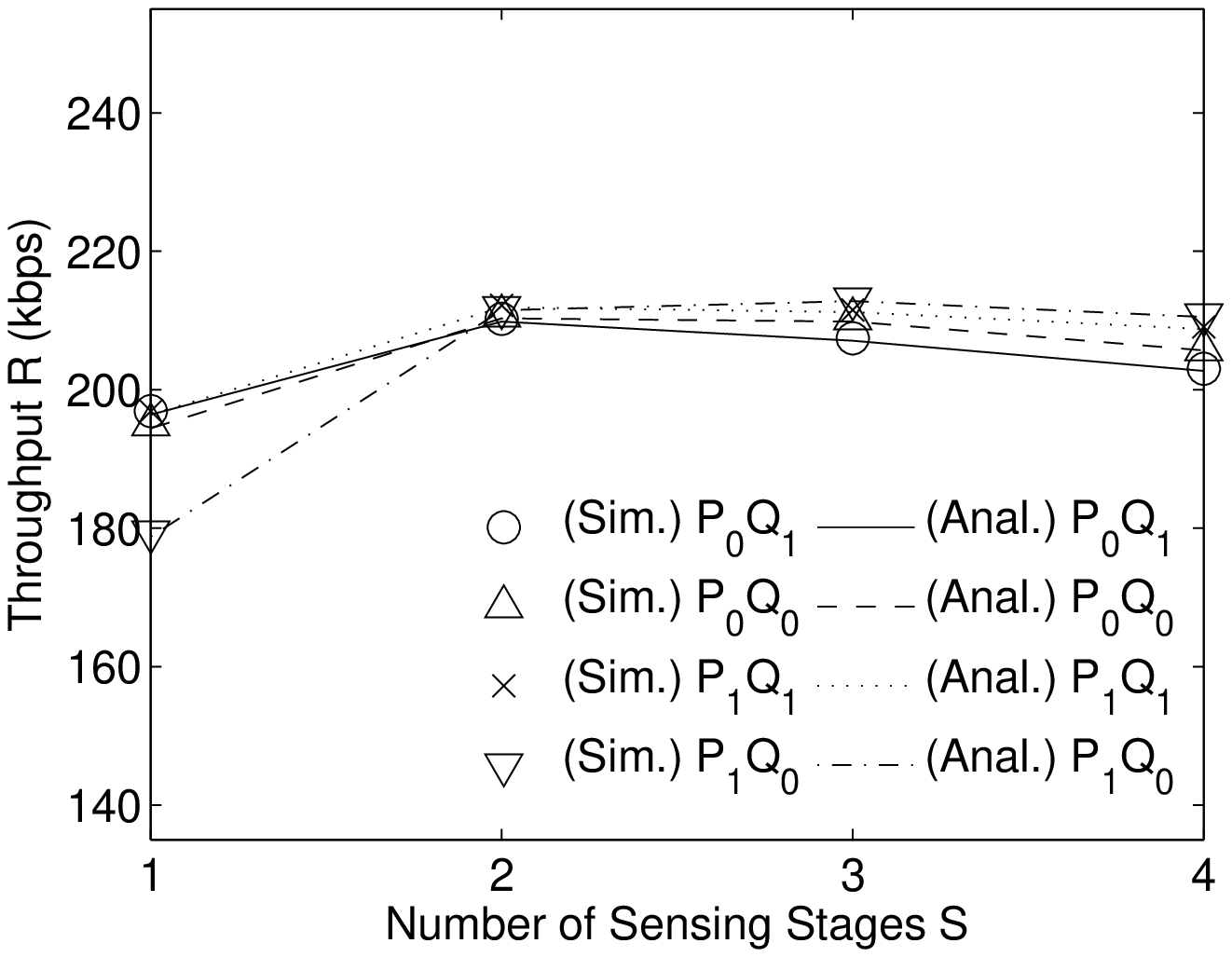}\label{fig:figs1a}}
\subfigure[]{\includegraphics[width=0.30\columnwidth]{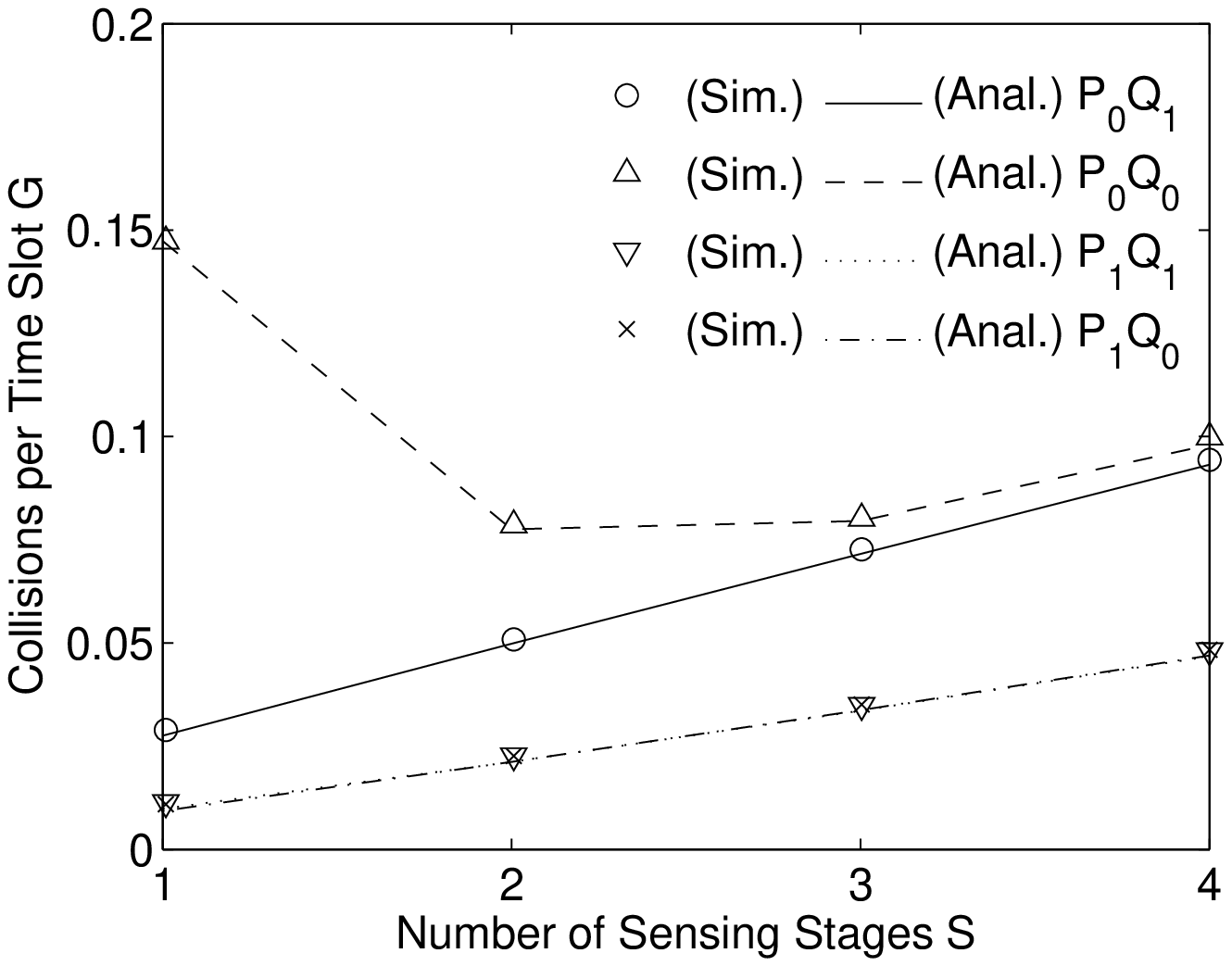}\label{fig:figs1b}}\\
\subfigure[]{\includegraphics[width=0.30\columnwidth]{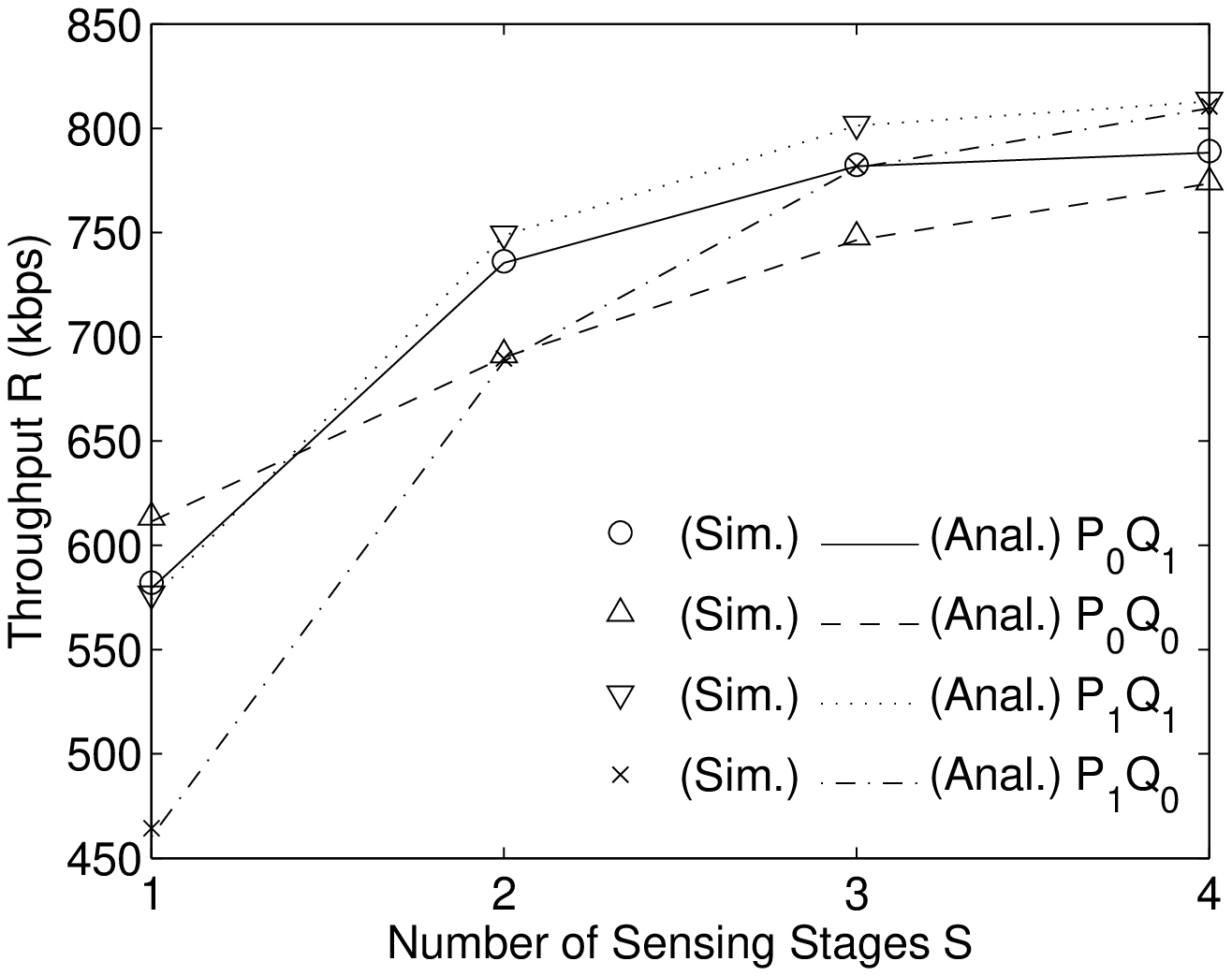}\label{fig:figs1c}}
\subfigure[]{\includegraphics[width=0.30\columnwidth]{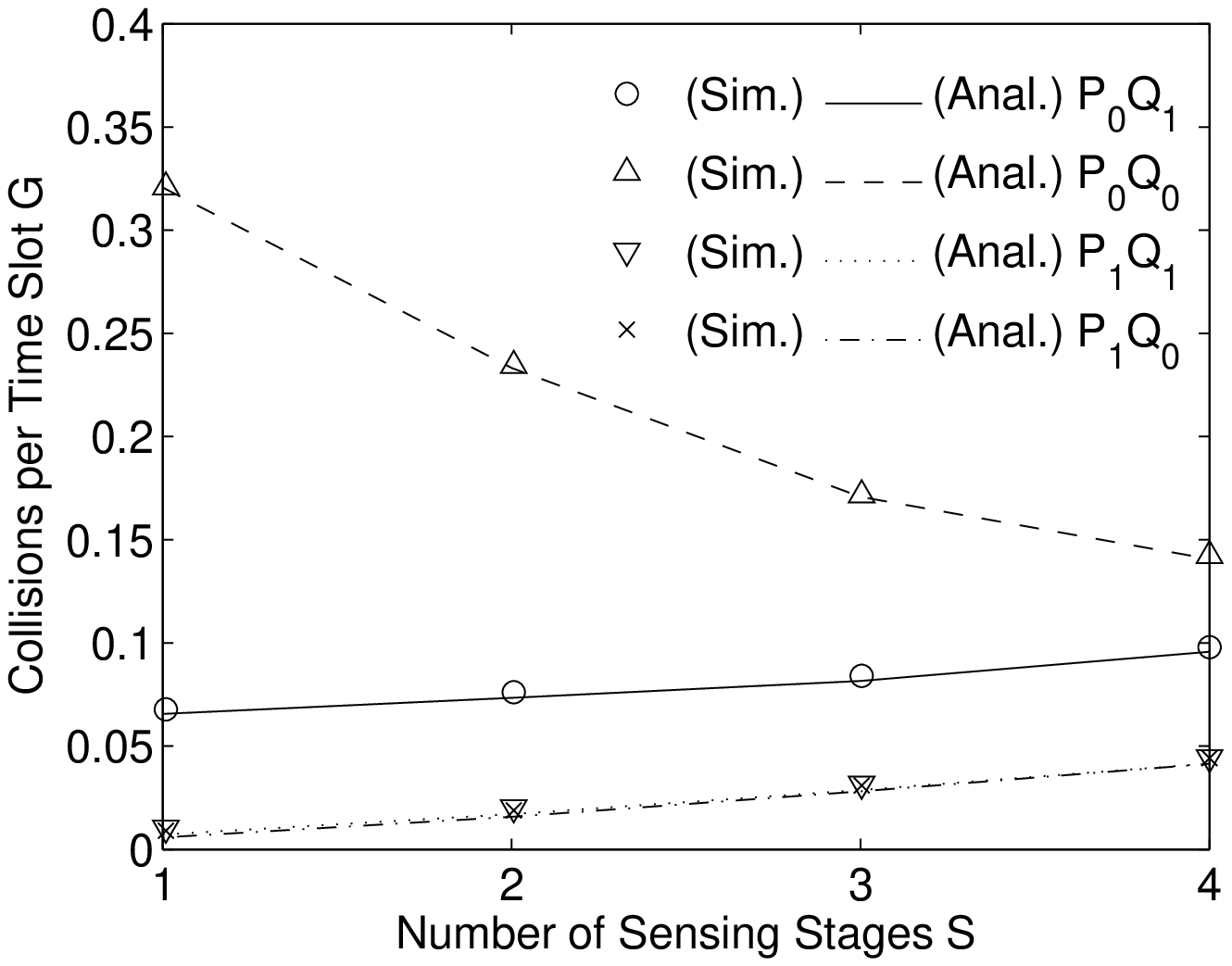}\label{fig:figs1d}}\\
\subfigure[]{\includegraphics[width=0.30\columnwidth]{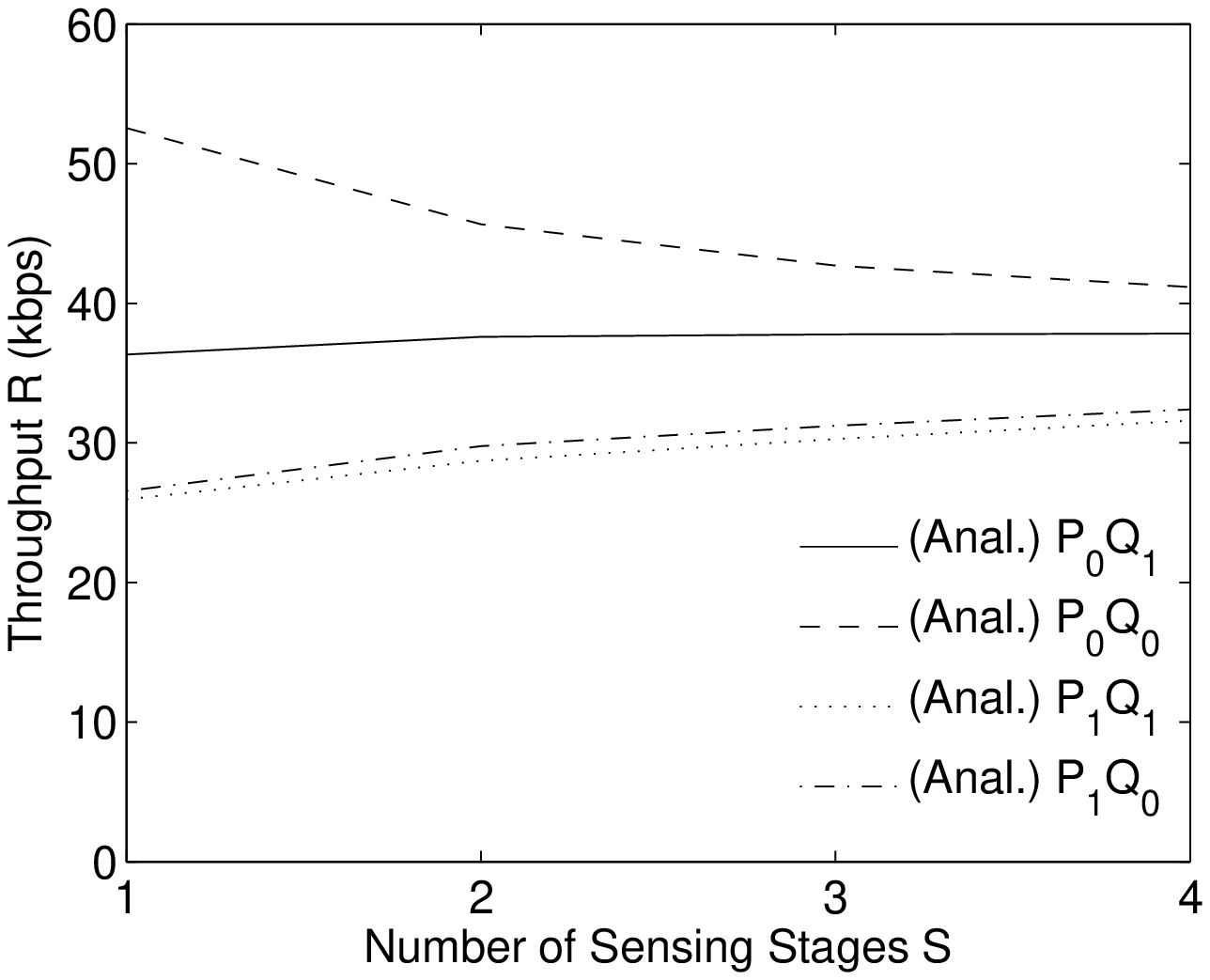}\label{fig:figs1e}}
\subfigure[]{\includegraphics[width=0.30\columnwidth]{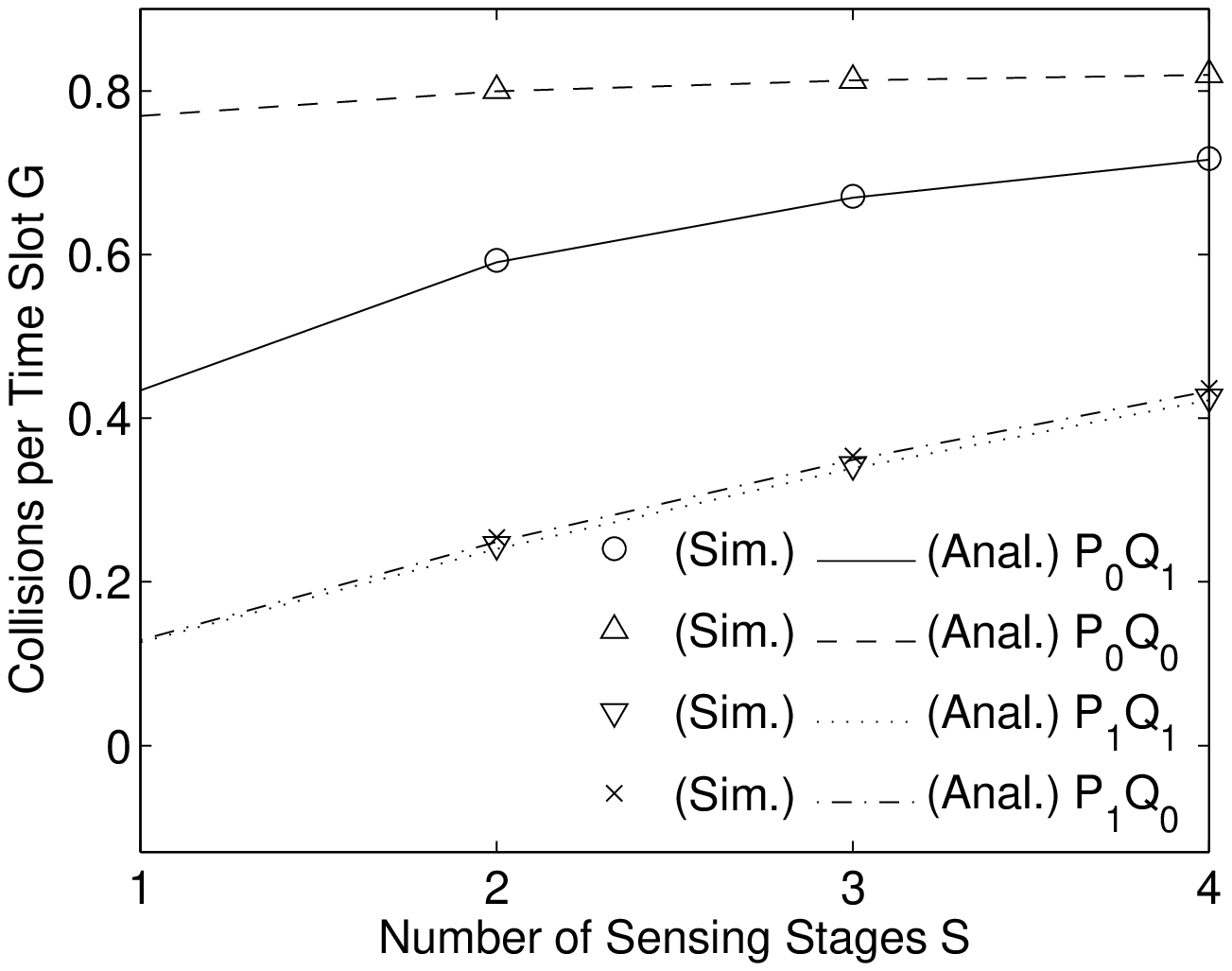}\label{fig:figs1f}}\\
\subfigure[]{\includegraphics[width=0.30\columnwidth]{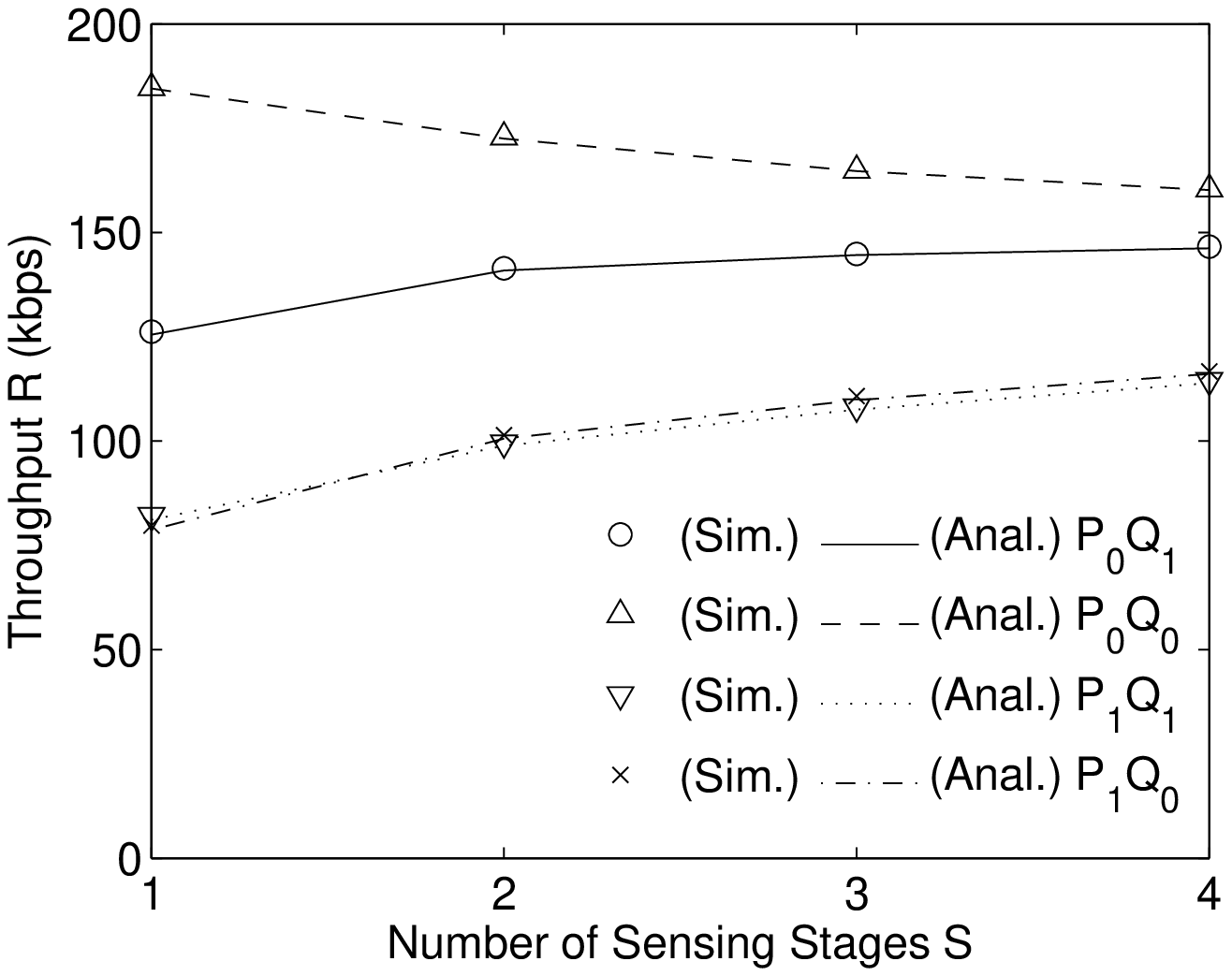}\label{fig:figs1g}}
\subfigure[]{\includegraphics[width=0.30\columnwidth]{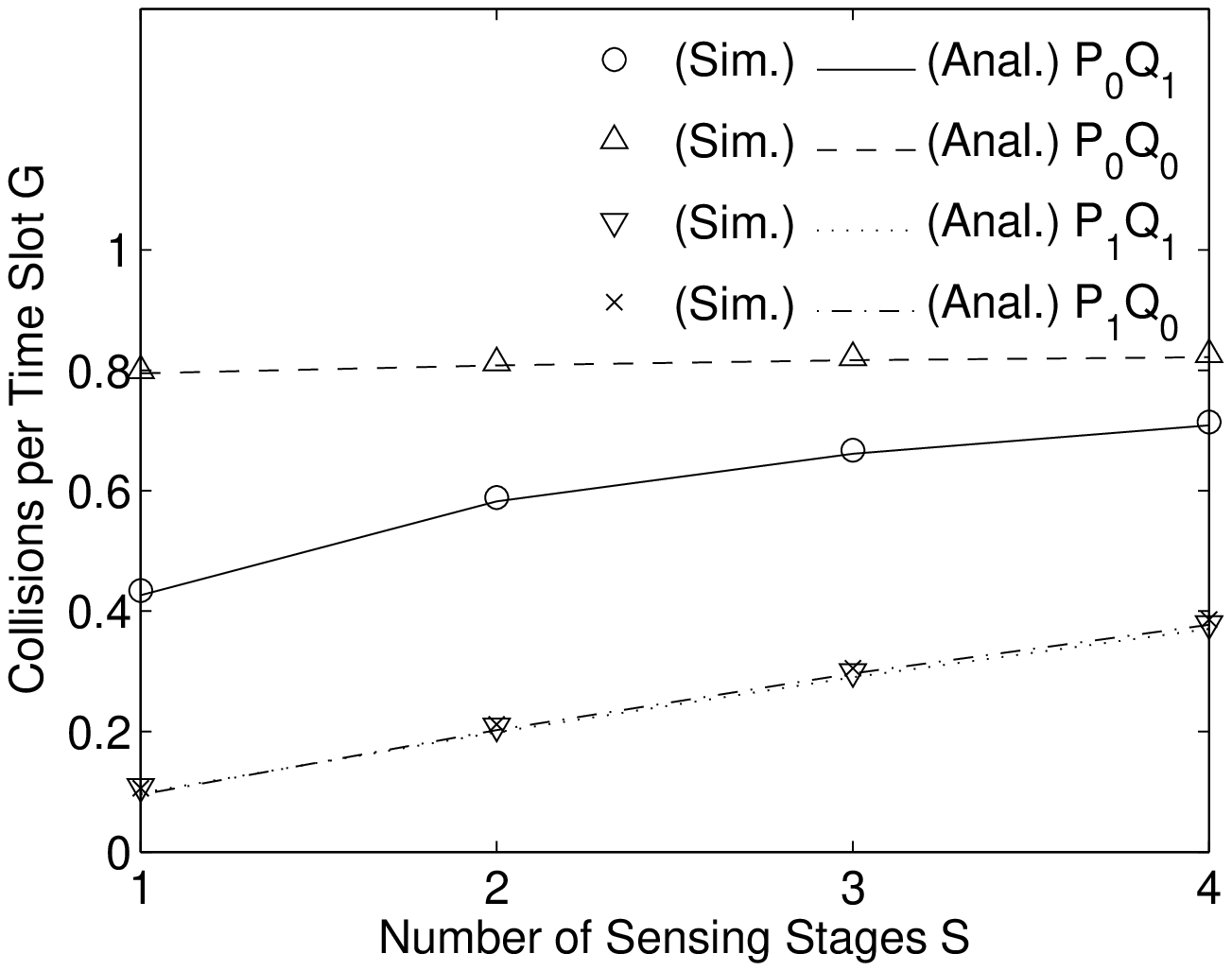}\label{fig:figs1h}}
\caption{$R$ and $G$ as functions of $S$ for the single narrowband radio case. (a) and (b) show $R$ and $G$, respectively, for slow PU traffic and long-$T_s$ low-$p_f$ sensing, (c) and (d) show $R$ and $G$, respectively, for slow PU traffic and short-$T_s$ high-$p_f$ sensing, (e) and (f) show $R$ and $G$, respectively, for fast PU traffic and long-$T_s$ low-$p_f$ sensing, and (g) and (h) show $R$ and $G$, respectively, for fast PU traffic and short-$T_s$ high-$p_f$ sensing. The slow and fast PU traffic scenarios and the long-$T_s$ low-$p_f$ and short-$T_s$ high-$p_f$ sensing techniques are as explained in Section~\ref{sec:sensing_stages_results}. Common parameters: $N=6$, $T=1\,$ms, $W=1$\,Mbps, $p_{s,a}=1$, $p_{s,d}=0$ and $B=0$.}
\label{fig:figs1}
\end{figure}

Results for the slow PU traffic model are shown in Fig.~\ref{fig:figs1a}, Fig.~\ref{fig:figs1b}, Fig.~\ref{fig:figs1c} and Fig.~\ref{fig:figs1d}. From our framework, it is apparent that the upper bound for $R$ equals $\left(1-\left(\frac{p_{p,a}}{p_{p,a}+ p_{p,d}}\right)^N\right)W$. For the slow PU traffic model, $R$ is upper bounded by 984.3\,kbps. For ideal sensing, that is, if $p_{f,s}=p_{f,t}=p_{m,s}=p_{m,t}=T_{s}=0$, $R$ for the four single narrowband radios algorithms approaches the upper bound. Using only one stage of ideal sensing results in $R$ that is less than 1\% below the upper bound. The loss in throughput is caused by the time spent by the SU in the quiet and pre-sensing modes (if they exist for the algorithm) and while transmitting on a channel occupied by a PU. However, for single stage non-ideal sensing, the yielded throughput for all algorithms is 33--39\% below the upper bound for the long-$T_s$ low-$p_f$ sensing technique and 38--53\% below the upper bound for the short-$T_s$ high-$p_f$ sensing technique. 

We now investigate the loss in $R$ for non-ideal sensing for the four algorithms. Considering P$_0$Q$_1$ and P$_1$Q$_0$ for $S=1$, it is clear that the quiet mode leads to a higher $R$ compared to the pre-sensing mode. This is because for P$_0$Q$_1$, a false alarm causes the SU to proceed from sensing stage one to the quiet mode. However, the quiet mode features a low probability of false alarm causing the SU to return to stage one of sensing, with high probability, and continue communicating on the vacant channel. On the other hand, for P$_1$Q$_0$, a single false alarm causes the SU to switch to a new channel that might not be vacant. P$_0$Q$_0$ which features no quiet or pre-sensing modes results in a higher $R$ than P$_1$Q$_0$ for $S=1$. Yet, this comes at the expense of collisions where $G$ for P$_0$Q$_0$ is 15 and 45 times more than that for P$_1$Q$_0$ for long-$T_s$ low-$p_f$ and short-$T_s$ high-$p_f$, respectively.

Considering multi-stage sensing with $S>1$ for slow PU traffic, as $S$ increases, the impact of false alarms on $R$ decreases. This is because more consecutive false alarms have to be generated in order to reach the quiet mode (for P$_0$Q$_1$ and P$_1$Q$_1$) or to switch to the next channel (for P$_0$Q$_0$ and P$_1$Q$_0$). On the other hand, increasing $S$ causes the SU to take a longer time, on average, to reach the quiet mode based on the successful detection of a PU. This causes $G$ to increase (for most algorithms) and consequently leads to a decrease in $R$. Thus, $S=2$ results in the highest $R$ for most algorithms for long-$T_s$ low-$p_f$. However, for short-$T_s$ high-$p_f$, $R$ increases with increasing $S$ due to the dominant effect of false alarms on all algorithms. In conclusion, for the slow PU traffic case, multi-stage sensing provides the flexibility to decrease the sensing time while increasing the resulting throughput. In our setup decreasing $T_s$ from $T_s = 0.24 T$ to $T_s = 0.1 T$ while increasing $S$ to $4$ stages leads to increasing $R$ by 14\%. Regarding the sensing algorithm, the pre-sensing mode results in a considerable decrease in $G$. Adding the quiet mode to the pre-sensing mode leads to an increase in $R$ for small $S$, and the gain in $R$ is more significant for higher probabilities of false alarm.

Results for the fast PU traffic case are shown in Fig.~\ref{fig:figs1e}, Fig.~\ref{fig:figs1f},  Fig.~\ref{fig:figs1g} and  Fig.~\ref{fig:figs1h}. As the PU traffic becomes faster, the probability that the SU switches to another channel increases. Hence, the stationary probability that the SU operates in the pre-sensing or quiet modes increases (for algorithms that have these stages) causing the resulting $R$ to decrease. For P$_0$Q$_1$, P$_1$Q$_1$ and P$_1$Q$_0$, increasing $S$ decreases the stationary probability of operating in the pre-sensing and quiet modes and accordingly, causes an increase in $R$. P$_0$Q$_0$ exhibits the highest $R$ for the fast PU traffic case, yet, this comes at the expense of collisions where $G$ is up to eight times higher than that for algorithms with pre-sensing. For P$_0$Q$_0$, increasing the number of sensing stages increases the time taken by the SUs to vacate the channel for an incoming PU, leading to an increase in $G$ and a decrease in $R$. Finally, note that the analytical results match perfectly with the Monte Carlo simulations. 

\subsection{Single Narrowband Radio: Throughput and Collisions versus Sensing Time Per Stage and Probability of False Alarm}
\label{sec:sensing_duration}

In this section, we analyze the effect of varying $T_s$ on throughput, $R$, and collisions, $G$, for the different sensing algorithms using the single narrowband radio architecture\footnote{More results on the effect of $T_s$ on the multi-channel multi-stage spectrum sensing algorithms in the context of energy consumption are presented in~\cite[Sec. IV-B]{gabran_submitted_2010}.}. For this experiment, in contrary to Section~\ref{sec:sensing_stages_results}, we set $S=2$ stages, $N = 3$ channels and $B = 2$ frames to explore the results from the perspective of a parameter set that is different from that of the previous section. Furthermore, we set $p_{p,a} = 0.01$, $p_{p,d} = 0.05$, $p_{s,a} = 0.1$ and we set the ratio $W\left(1-\frac{T_s}{T}\right)\frac{p_{s,a}}{p_{s,a}+p_{s,d}}$ (the average generated SU traffic) to 500\,kbps. For the different values of $T_s=[50,500]$\,$\mu$s, the energy detection sensing threshold is chosen to keep $p_{m,s}$ constant at $0.1$, while $p_{f,s}=[0.23,0.013]$. The unsuccessful frame delivery rate is non-zero as frames are lost either because of collisions with PU frames or buffer overflow when there are no channel vacancies. For clarity of presentation, we assume a QoS constraint where the maximum unsuccessful frame delivery rate is set to $0.1$ and we do not present the results that do not satisfy the QoS constraint.
\begin{figure}
\centering
\subfigure[]{\includegraphics[width=0.49\columnwidth]{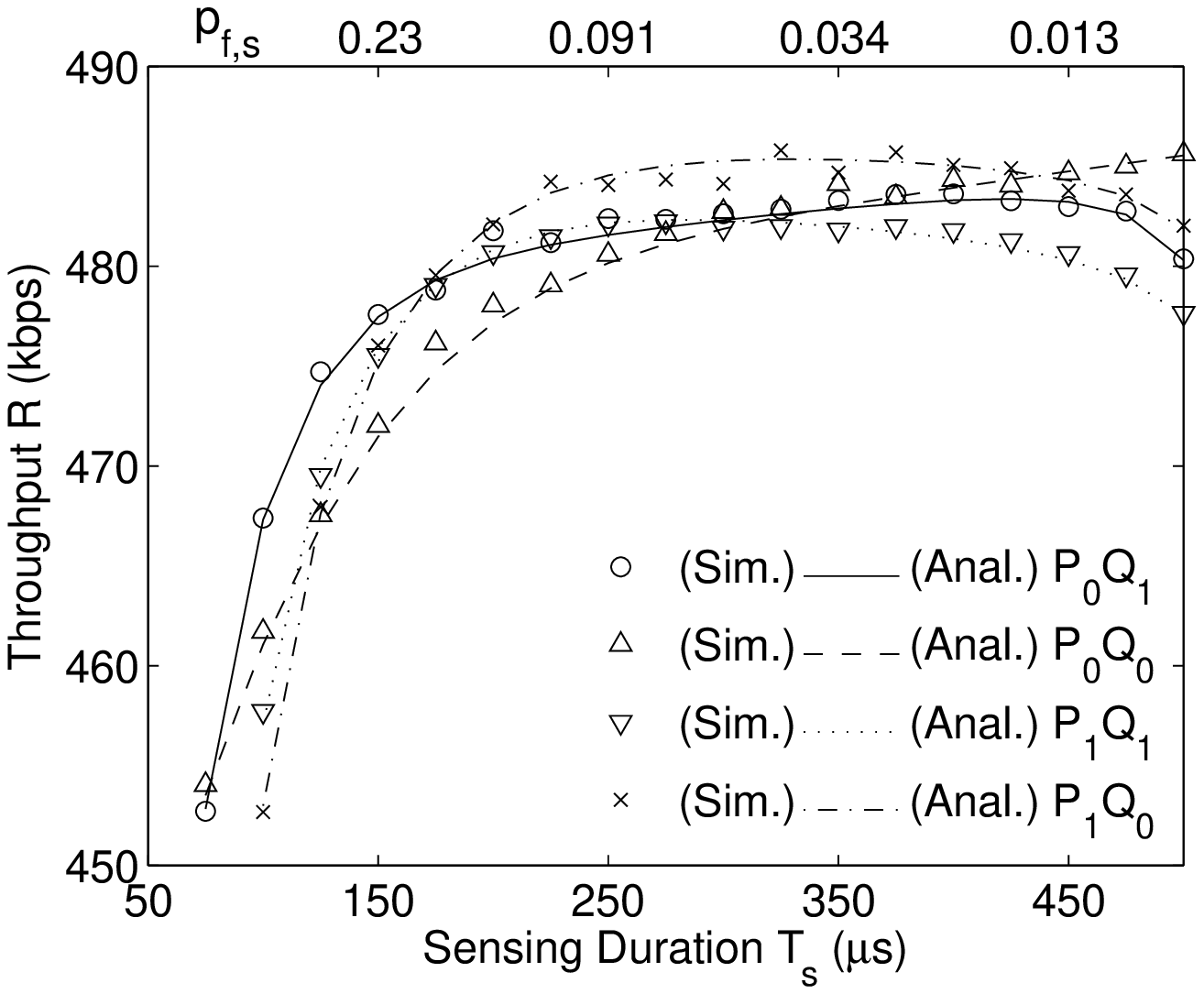}\label{fig:sensing_phase1}}
\subfigure[]{\includegraphics[width=0.49\columnwidth]{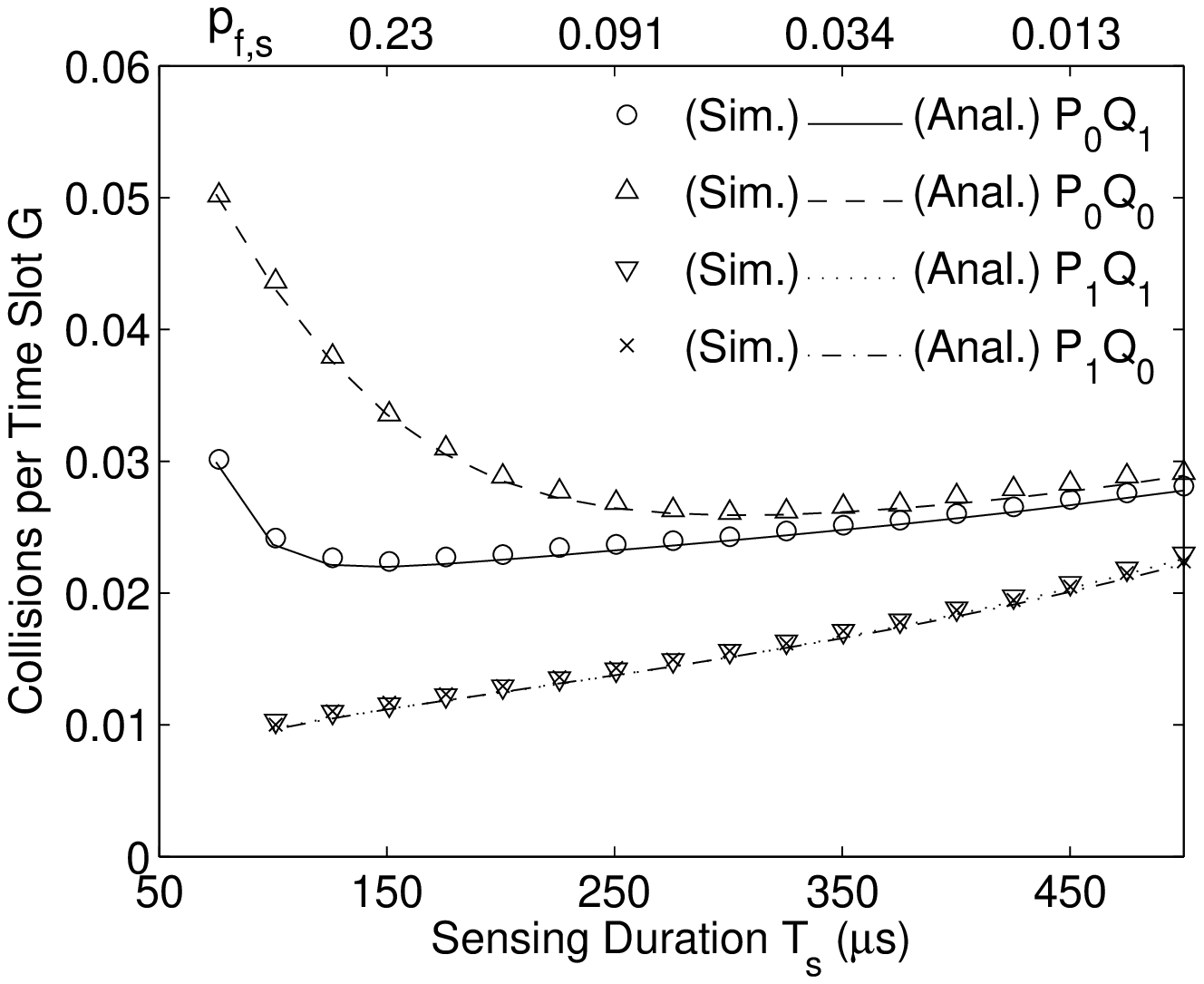}\label{fig:sensing_phase2}}
\caption{Single narrowband radios case: (a) throughput $R$ and (b) collision probability $G$ as functions of $T_s$. Common parametrs: $T=1\,$ms, $S = 2$, $N = 3$, $B=2$, $p_{m,s} = 0.1$, $p_{p,a} = 0.01$, $p_{p,d} = 0.05$, $p_{s,a}=0.1$ and the generated throughput equals $500\,$kbps. Note the respective probability of false alarm, $p_{f,s}$, for each sensing time are marked at the top of each figure.}
\label{fig:sensing_phase}
\end{figure}

The impact of varying $T_s$ on $R$ is shown in Fig.~\ref{fig:sensing_phase1}. Results show that for P$_0$Q$_1$, P$_1$Q$_1$ and P$_1$Q$_0$, there is an optimal value of $T_s$ that maximizes throughput (resulting from the sensing/throughput tradeoff~\cite{liang_twc_2008,peh_tvt_2009}) while for P$_0$Q$_0$, $R$ increases monotonically with $T_s$. This is because increasing $T_s$ decreases the probability of false alarm leading to an increase in the utilized spectral vacancy and hence, an increase in throughput. On the other hand, increasing $T_s$ increases the number of transmitted frames needed to meet the throughput requirement. As a result, for long $T_s$, increasing $T_s$ increases the probability of buffer overflow causing a decrease in throughput. However, this is not the case for P$_0$Q$_0$ as frames do not get buffered, as explained in Section~\ref{sec:single_radio_example}.

The relationship between $T_s$ and $G$ is presented in Fig.~\ref{fig:sensing_phase2}. Increasing $T_s$ increases the number of transmitted frames, to maintain generated throughput at 500\,kbps, and thus causes an increase in the collision probability. However, shorter $T_s$ results in a higher false alarm probability. This causes the SU to switch from a vacant channel to a new channel that might be occupied by a PU resulting in a collision. The pre-sensing stage decreases collisions caused by switching to an occupied channel. Accordingly, for P$_1$Q$_1$ and P$_1$Q$_0$, $G$ increases monotonically with $T_s$, while on the other hand, the relationship between $T_s$ and $G$ is convex and non-monotonic for $T_s$ for P$_1$Q$_1$ and P$_1$Q$_0$. Comparing Fig.~\ref{fig:sensing_phase1} and Fig.~\ref{fig:sensing_phase2} it is clear that the best protocol option (that maximizes throughput, while at the same time minimizes collisions) is P$_1$Q$_0$.

\subsection{Single Narrowband Radio: Throughput versus Buffer Size}
\label{sec:buffer_size}

For this experiment, we set $S=2$ stages, $N = 3$ channels, $p_{m,s}=0.1$ and $p_{p,a} = p_{p,d} = 0.01$. We analyze the two scenarios considered earlier in Section~\ref{sec:sensing_stages_results}: (i) one with slow SU traffic (relative to the slot duration) with $p_{s,a} = p_{s,d} = 0.01$ and the long-$T_s$ low-$p_f$ sensing technique, and (ii) one with fast PU traffic with $p_{s,a} = 0.5$ and $p_{s,d} = 0.1$ and the short-$T_s$ high-$p_f$ sensing technique.

\begin{figure}
\centering
\subfigure[]{\includegraphics[width=0.49\columnwidth]{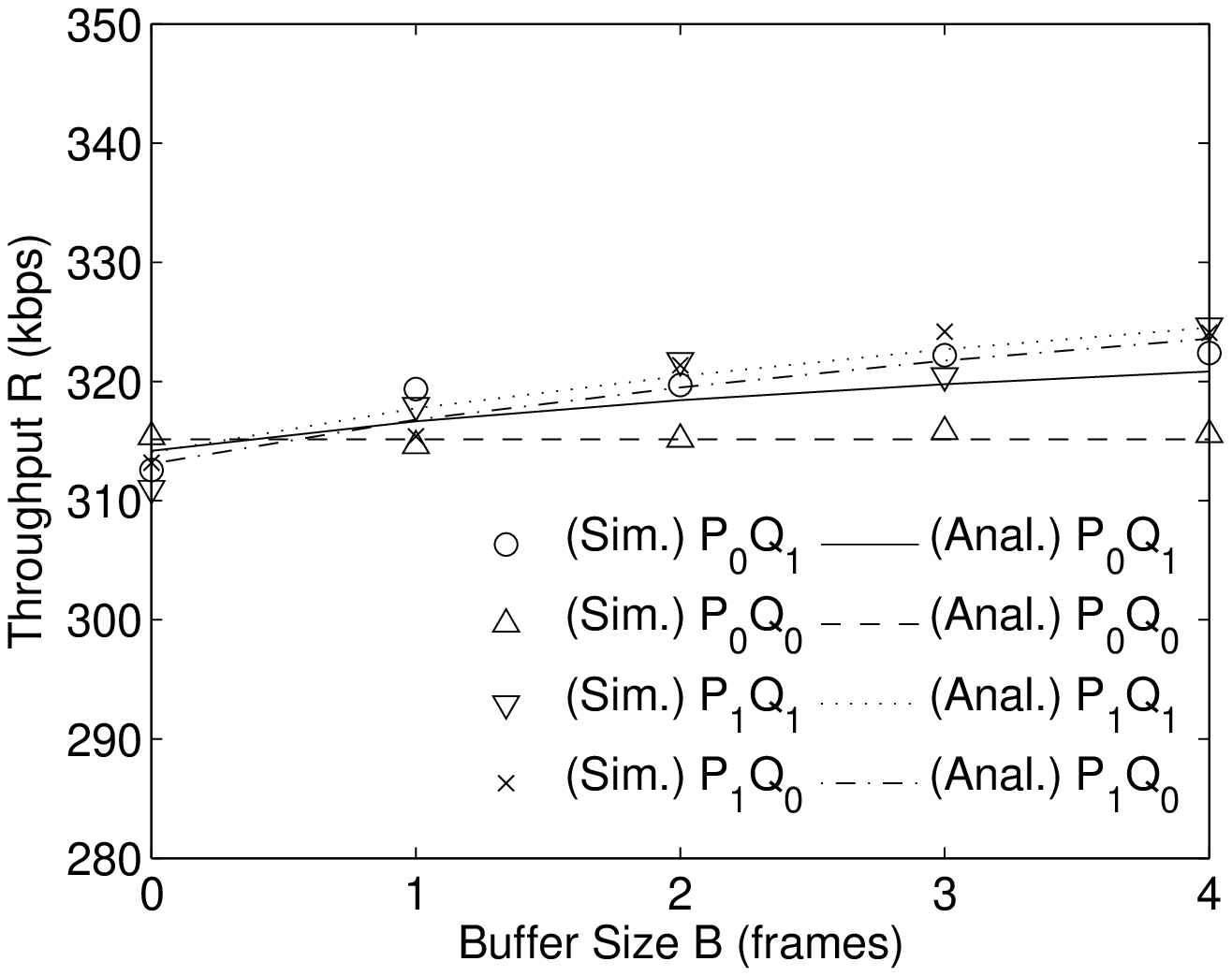}\label{fig:figs3a}}
\subfigure[]{\includegraphics[width=0.49\columnwidth]{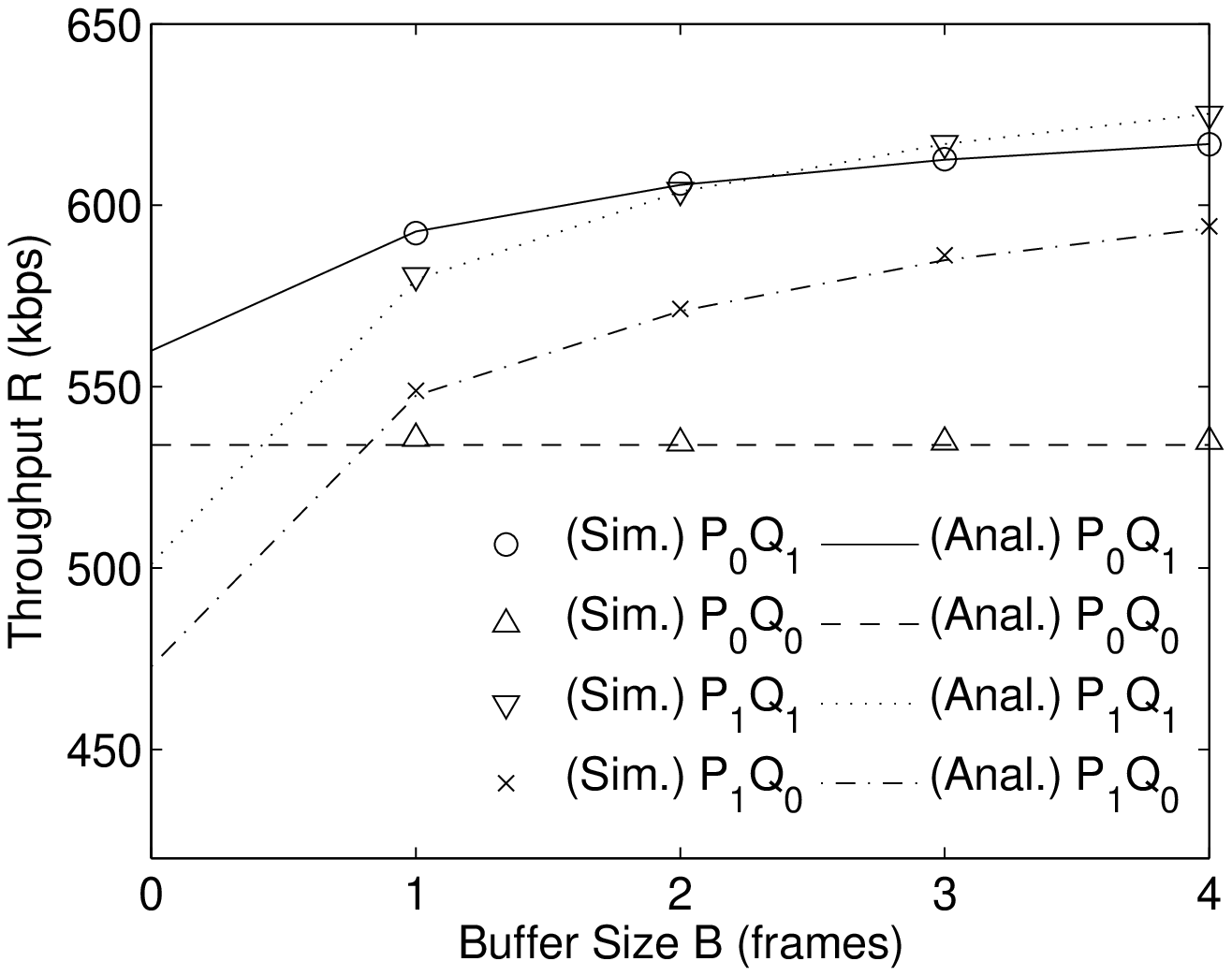}\label{fig:figs3b}}
\caption{$R$ as a function of $B$ for the single narrowband radio case for (a) slow SU traffic and long-$T_s$ low-$p_f$ sensing and (b) fast SU traffic and short-$T_s$ high-$p_f$ sensing. The slow and fast SU traffic scenarios are as explained in Section~\ref{sec:buffer_size} and the long-$T_s$ low-$p_f$ and short-$T_s$ high-$p_f$ sensing techniques are as explained in Section~\ref{sec:sensing_stages_results}. Moreover: $S=2$, $N=3$, $T=1\,$ms, $W=1$\,Mbps, $p_{p,a}=p_{p,d}=0.01$.}
\label{fig:figs3}
\end{figure}

The impact of increasing $B$ on $R$ is shown in Fig.~\ref{fig:figs3}. Frames are buffered whenever the radio is in the quiet or pre-sensing modes. Accordingly, no frames are buffered for P$_0$Q$_0$, thus, $R$ is independent of $B$. The buffered frames are transmitted whenever no new SU frames are generated and the SU buffer is not empty. For P$_0$Q$_1$, P$_1$Q$_0$ and P$_1$Q$_1$, increasing $B$ increases the probability that the new generated frame will not be discarded due to the absence of an idle channel. Increasing $B$ increases $R$ where the impact of $B$ on $R$ is less significant for large values of $B$. Note, again, the close match between the analysis and simulations.

\subsection{Parallel Narrowband Radios: Throughput and Collisions versus Number of Stages}
\label{sec:wideband_stages}

\begin{figure}
\centering
\subfigure[]{\includegraphics[width=0.49\columnwidth]{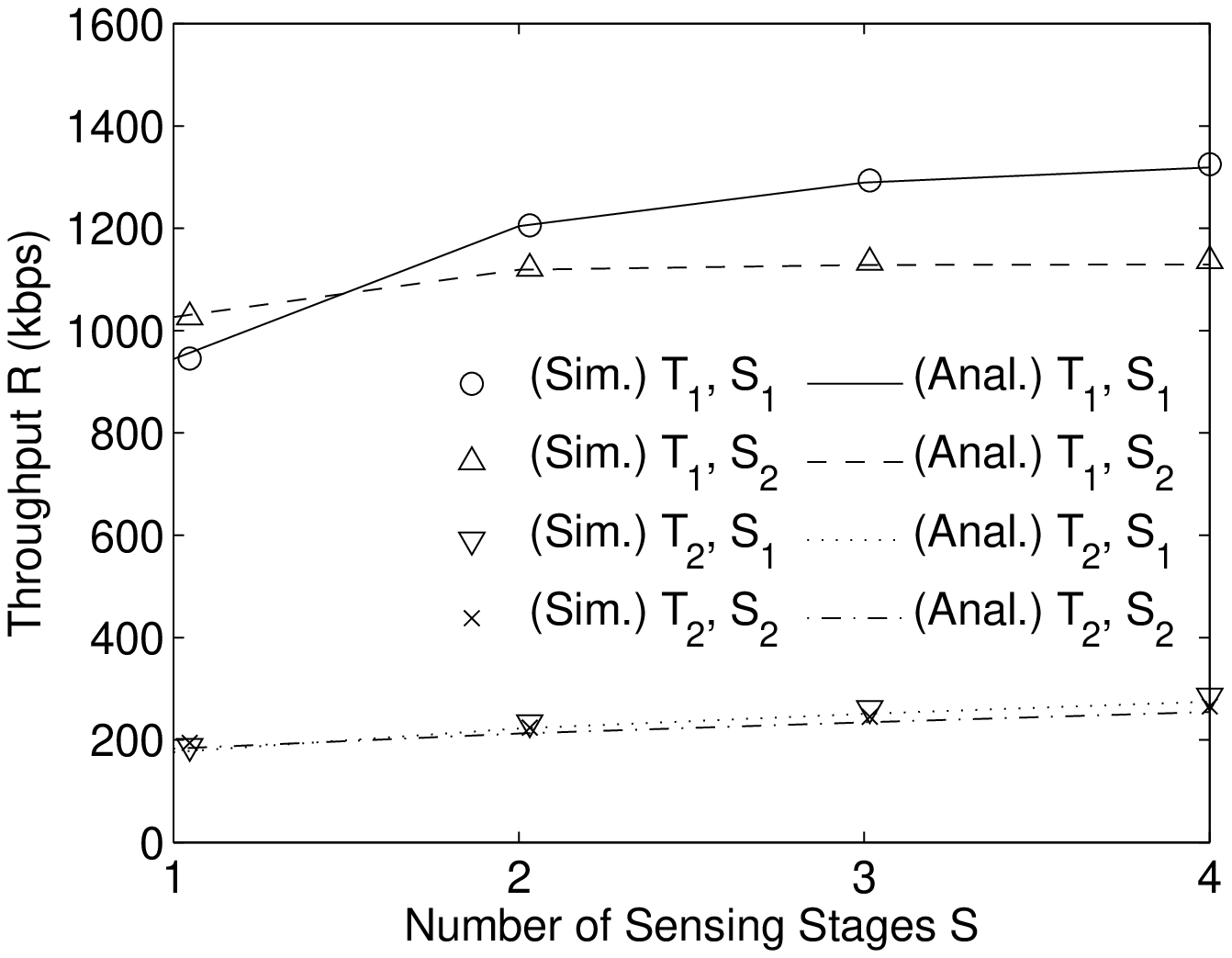}\label{fig:figs_wb_1}}
\subfigure[]{\includegraphics[width=0.49\columnwidth]{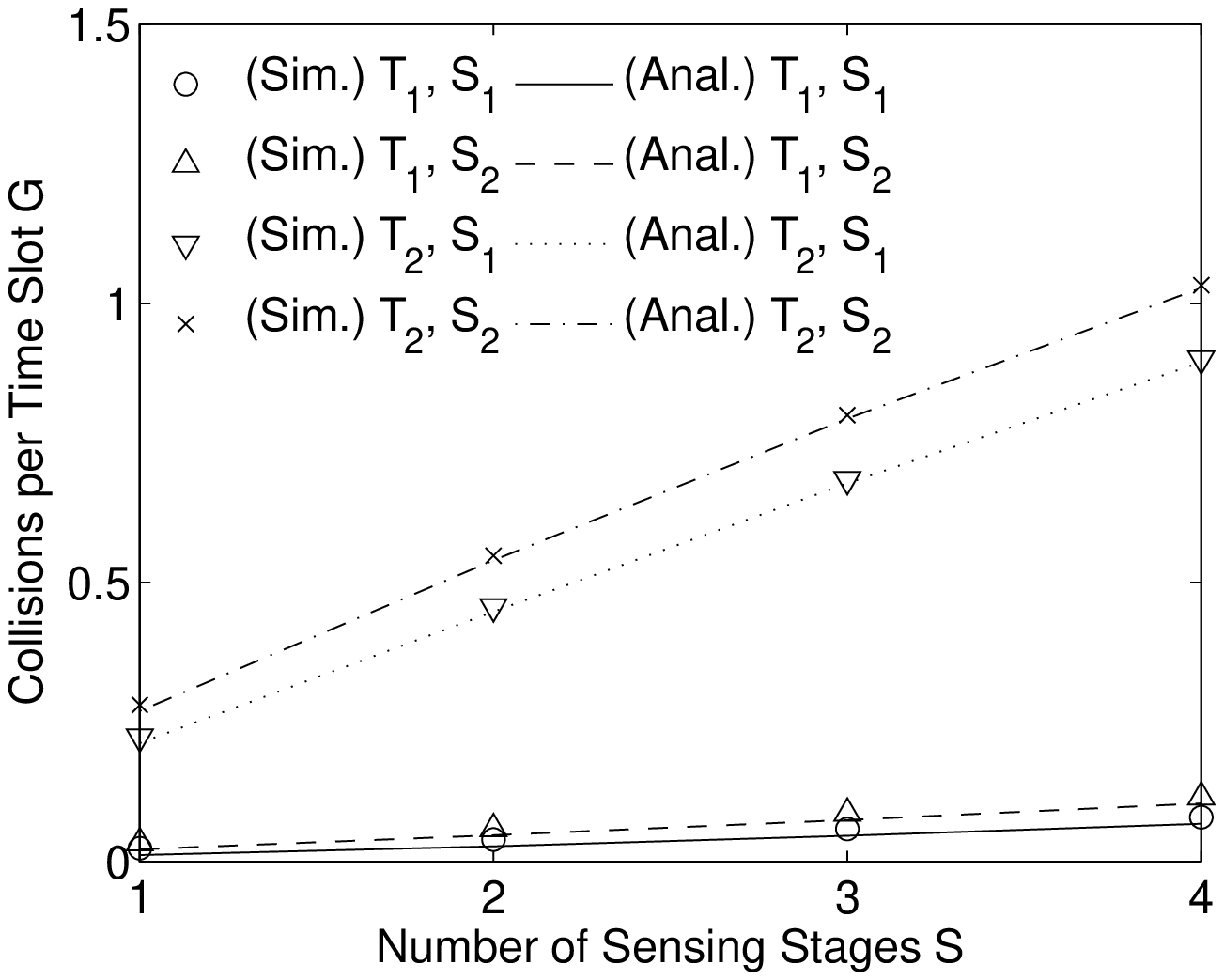}\label{fig:figs_wb_2}}
\caption{Parallel narrowband radios case: (a) throughput $R$ and (b) collision probbaility $G$ as functions of $S$. $T_1$ and $T_2$ denote slow and fast PU traffic, respectively, and $S_1$ and $S_2$ denote short-$T_s$ high-$p_f$ and long-$T_s$ low-$p_f$ sensing techniques, respectively. The slow and fast PU traffic scenarios and the long-$T_s$ low-$p_f$ and short-$T_s$ high-$p_f$ sensing techniques are as explained in Section~\ref{sec:sensing_stages_results}. Moreover: $N=3$, $T=1\,$ms, $W=1$\,Mbps, $p_{s,a}=1$, $p_{s,d}=0$, $B=0$. Note that $G$ exceeds one as, in this case, it represents the expected number of collisions for $N=3$ per time slot.}
\label{fig:figs_wb}
\end{figure}

We investigate the effect of varying $S$ on $R$ and $G$ for parallel narrowband radios. We use the same set of parameters as in Section~\ref{sec:sensing_stages_results}, except for $N = 3$ channels. We consider the two PU traffic scenarios analyzed earlier: the slow PU traffic with $p_{s,a} = p_{s,d} = 0.01$ and the fast PU traffic with $p_{s,a} = 0.5$ and $p_{s,d} = 0.1$. We also consider the two sensing options explored earlier: the long-$T_s$ low-$p_f$ and the short-$T_s$ high-$p_f$ sensing techniques. We analyze the scenarios with the four permutations of PU traffic and sensing options. The impact of varying $S$ on $R$ and $G$ is shown in Fig.~\ref{fig:figs_wb_1} and Fig.~\ref{fig:figs_wb_2}, respectively. 

Regarding $R$, increasing $S$ increases $R$ for the four scenarios. This is because more consecutive false alarms have to be generated in order for a radio to switch to the quiet mode while the channel is vacant. Moreover, the results confirm the intuition that $R$ is directly proportional to $N$ for the tested scenarios. Similarly, increasing $S$ increases $G$ for the four scenarios. This is because increasing $S$ increases the average time taken by the SU to switch to the quiet mode when a PU arrives. Finally, results clearly show the tradeoff between $T_s$, $R$ and $G$, and the improvement in $R$ and $G$ caused by employing multi-stage sensing. This is demonstrated in the slow PU traffic case with $S = 4$, where $R$ is higher by 17\% and $G$ is lower by 35\% for scenarios employing the short-$T_s$ high-$p_f$ sensing technique compared to those employing the long-$T_s$ low-$p_f$ sensing technique. As in the previous cases, analysis matches perfectly with the simulations.

\subsection{Single versus Parallel Narrowband Radios: Throughput and Collisions versus the Number of Available Channels}
\label{sec:comparison}

\begin{figure}
\centering
\subfigure[]{\includegraphics[width=0.49\columnwidth]{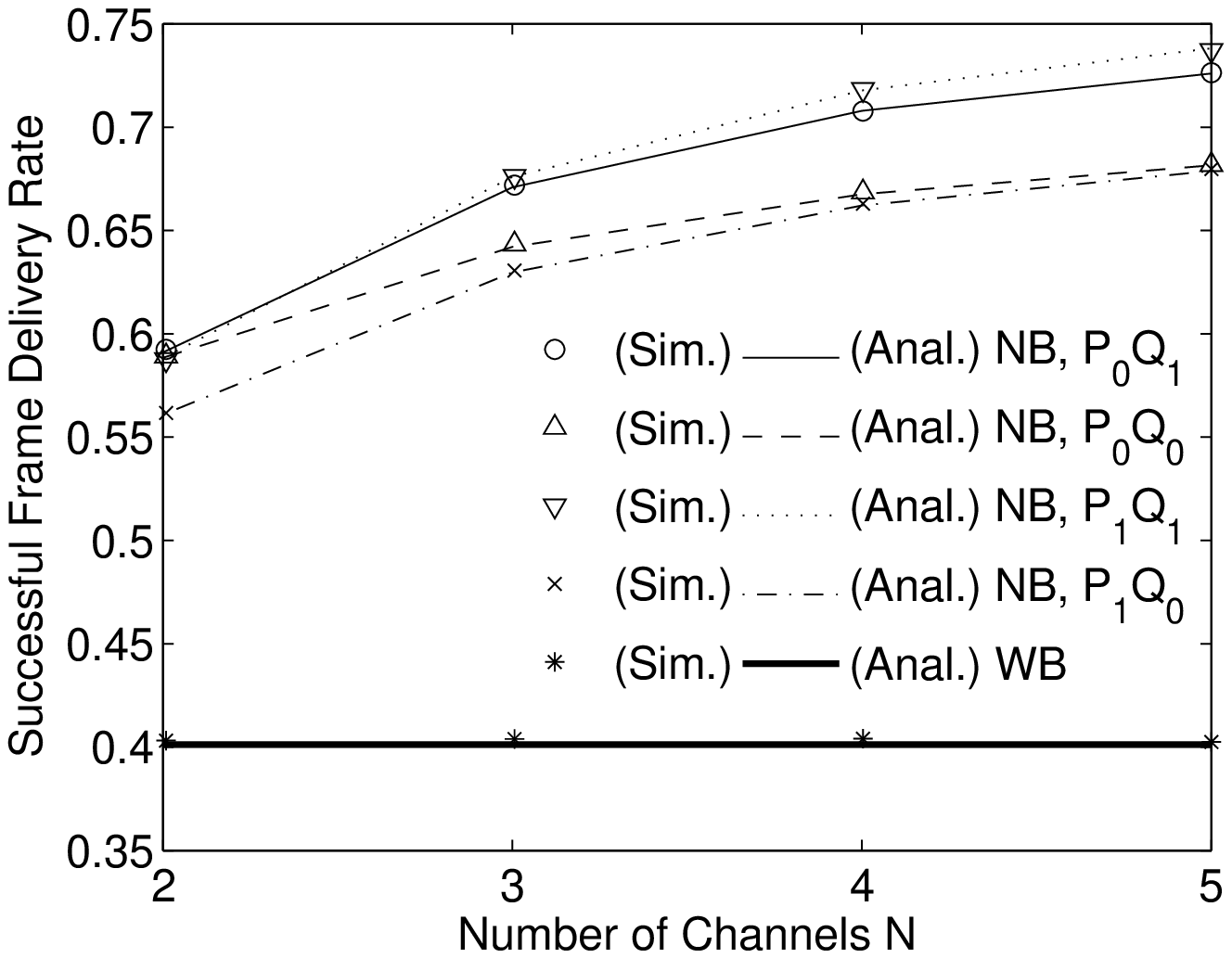}\label{fig:narrow_wide1}}
\subfigure[]{\includegraphics[width=0.49\columnwidth]{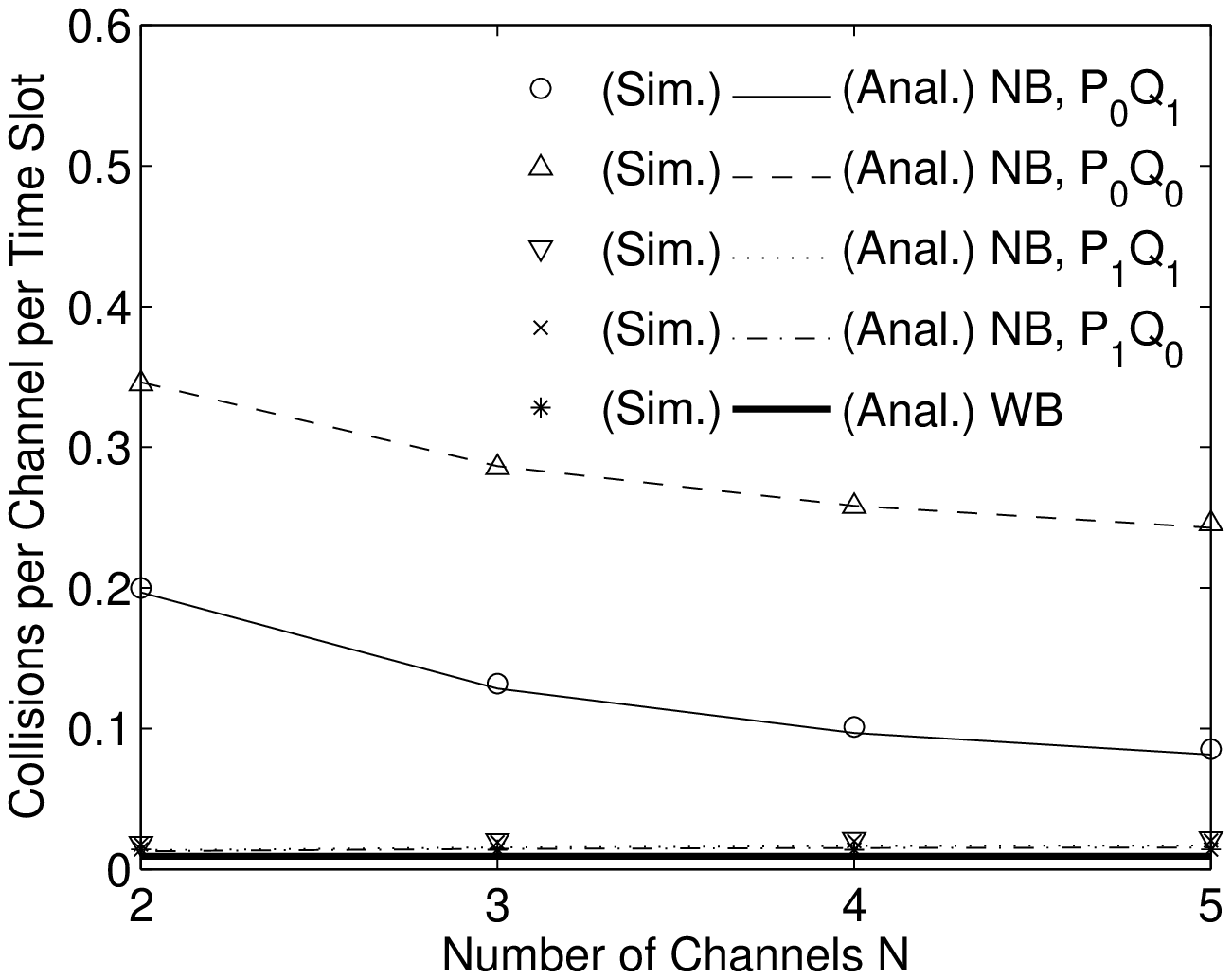}\label{fig:narrow_wide2}}
\caption{Performance comparison of single and parallel narrowband radios: (a) successful frame delivery rate and (b) probability of collisions per channel per time slot as functions of $N$. Common parameters: $T=1\,$ms, $S = 2$, $p_{s,a}=1$, $p_{s,d}=0$, $B=0$. The short-$T_s$ high-$p_f$ sensing technique described in Section~\ref{sec:sensing_stages_results} is adopted; NB: single narrowband radios, WB: parallel narrowband radios.}
\label{fig:narrow_wide}
\end{figure}

Finally, in this section we present the relationship between the performance metrics, $R$ and $G$, and the number of available PU channels, $N$. At the same time, we compare the performance of single and parallel narrowband radios. For this experiment, we set $S=2$ stages, $B=0$ frames, $p_{m,s}=1$ and $p_{p,a} = p_{p,d} = 0.01$. As an example we adopt the short-$T_s$ high-$p_f$ sensing technique described in Section~\ref{sec:sensing_stages_results} with $T_s = 0.1 T$. We consider saturated SU traffic ($p_{s,a} = 1$ and $p_{s,d} = 0$) for both single and parallel narrowband radios. That is, the single narrowband radio generates one frame per time slot, as opposed to $N$ frames per time slot for the parallel narrowband radios case. As the throughput of the parallel narrowband radios is $N$ times higher than that of the single narrowband radio, the resulting number of collisions per time slot is normalized by $N$ to achieve a fair comparison. Moreover, instead of using the throughput metric, we introduce a new metric denoted as the successful frame delivery rate, which is defined as $R/(NW)$, i.e. the ratio between the successfully transmitted throughput and the generated throughput (which in this case equals to the available channel capacity). The number of PU channels, $N$, is increased from two to five and the resulting metrics are presented in Fig.~\ref{fig:narrow_wide}.

The results show that the successful frame delivery rate for all algorithms of single narrowband radios is higher than that of parallel narrowband radios. We want to stress, however, that the total (unnormalized) throughput for parallel wideband radios (not shown due to space constraints) is higher than that for single narrowband radios. Moreover, the successful frame delivery rate increases with $N$ for single narrowband radios but is constant relative to $N$ for parallel narrowband radios. This is because for single narrowband radios, increasing the available channels increases the probability of the SU finding a vacant channel to occupy. Note that, the successful frame delivery rate is upper bounded by $1-\left(\frac{p_{p,a}}{p_{p,a}+ p_{p,d}}\right)^N$. On the other hand, regarding the parallel narrowband radios case, as we assume saturated traffic, $N$ frames are generated by the $N$ radios every time slot, hence, the radios can be regarded as operating independently. As a result, increasing the number of available channels has no impact on the successful frame delivery rate.

Regarding the normalized probability of collisions per time slot, for parallel narrowband radios it is slightly less than that for single narrowband radios employing the pre-sensing stage, but significantly less than that for single narrowband radios with no pre-sensing. Furthermore, the normalized probability of collisions per time slot for parallel narrowband radios do not vary with $N$ due to the independence between the parallel radios caused by traffic saturation. Finally, the normalized probability of collisions per time slot decreases significantly with $N$ for single narrowband radios with no pre-sensing as increasing the available channels increases the probability of the SU operating on a vacant channel, instead of switching to a new channel that might be occupied by a PU.

\section{Conclusions}
\label{sec:conclusions}

We present a unified analytical framework, based on Markov chain analysis, for evaluating multi-channel multi-stage spectrum sensing algorithms for Opportunistic Spectrum Access networks. Multi-stage sensing provides more flexibility in optimizing the systemÕs performance by varying the sensing time per stage and the number of sensing stages. Therefore, in the model we consider a variety of algorithm design examples that feature prolonged sensing stages before accessing or leaving a channel. We also consider the temporal variation in the activity of the primary users as well as the secondary users. Furthermore, we analyze operation on multiple channels for nodes with single and parallel narrowband radios.

The results demonstrate the tradeoff inherent to multi-stage spectrum sensing between the secondary user throughput and the collision probability between primary and secondary users. In most cases, increasing the number of sensing stages increases both the throughput and the collision probability. For example, for a single narrowband radio, increasing the number of sensing stages from one to four stages, for a slow primary user traffic scenario, results in an increase in throughput of 36\%, yet causes an increase in the collision probability by 46\%. This tradeoff applies for both single and parallel narrowband radio systems. Moreover, the optimal number of sensing stages has a high dependence on the primary user traffic. Generally, having more than two stages does not result in a significant change in throughput yet causes an increase in collisions.

Regarding the sensing algorithms, the pre-sensing mode always results in a decrease in the expected number of collisions, while the quiet mode causes an increase in throughput in most scenarios. Results also show that increasing the secondary userÕs buffer size results in an increase in throughput, and the increase is more significant for fast secondary user traffic. Finally, comparing single and parallel narrowband radio systems, single narrowband radios result in a higher successful frame delivery rate as opposed to parallel narrowband radios. However, this comes at the expense of the maximum achievable throughput which scales with the number of available channels for parallel narrowband radios.


\begin{thebibliography}{10}
\providecommand{\url}[1]{#1}
\csname url@rmstyle\endcsname
\providecommand{\newblock}{\relax}
\providecommand{\bibinfo}[2]{#2}
\providecommand\BIBentrySTDinterwordspacing{\spaceskip=0pt\relax}
\providecommand\BIBentryALTinterwordstretchfactor{4}
\providecommand\BIBentryALTinterwordspacing{\spaceskip=\fontdimen2\font plus
\BIBentryALTinterwordstretchfactor\fontdimen3\font minus
  \fontdimen4\font\relax}
\providecommand\BIBforeignlanguage[2]{{%
\expandafter\ifx\csname l@#1\endcsname\relax
\else
\language=\csname l@#1\endcsname
\fi
#2}}

\bibitem{gabran_submitted_2010}
W.~{Gabran}, P.~{Pawe{\l}czak}, and D.~{\v{C}abri{\'c}}, ``Multi-channel
  multi-stage spectrum sensing: Link layer performance and energy
  consumption,'' in \emph{Proc. IEEE DySPAN 2011}, Aachen, Germany, EU, May
  3--6, 2011.

\bibitem{staple_spectrum_2004}
G.~{Staple} and K.~{Werbach}, ``The end of spectrum scarcity,'' \emph{{IEEE}
  Spectr.}, vol.~41, no.~3, pp. 48--52, Mar. 2004.

\bibitem{Zhao_sigprocmag_2007}
Q.~{Zhao} and B.~M. {Sadler}, ``A survey of dynamic spectrum access: Signal
  processing, networking, and regulatory policy,'' \emph{{IEEE} Signal
  Processing Mag.}, vol.~24, no.~3, pp. 79--89, May 2007.

\bibitem{yucek_commsurv_2009}
T.~{Y{\"u}cek} and H.~{Arslan}, ``A survey of spectrum sensing algorithms for
  cognitive radio applications,'' \emph{{IEEE} Commun. Surv. Tutor.}, vol.~11,
  no.~1, pp. 116--130, First Quarter 2009.

\bibitem{liang_twc_2008}
Y.-C. {Liang}, Y.~{Zeng}, E.~C. {Peh}, and A.~T. {Hoang}, ``Sensing-throughput
  tradeoff for cognitive radio networks,'' \emph{{IEEE} Trans. Wireless
  Commun.}, vol.~7, no.~4, pp. 1326--1337, Apr. 2008.

\bibitem{peh_tvt_2009}
E.~C.~Y. {Peh}, Y.-C. {Liang}, Y.~L. {Guan}, and Y.~{Zeng}, ``Optimization of
  cooperative sensing in cognitive radio networks: A sensing-throughput
  tradeoff view,'' \emph{{IEEE} Trans. Veh. Technol.}, vol.~58, no.~9, pp.
  5294--5299, Nov. 2009.

\bibitem{pawelczak_tvt_2009}
P.~Pawe{\l}czak, S.~{Pollin}, H.-S.~W. {So}, A.~{Bahai}, R.~V. {Prasad}, and
  R.~{Hekmat}, ``Performance analysis of multichannel medium access control
  algorithms for opportunistic spectrum access,'' \emph{{IEEE} Trans. Veh.
  Technol.}, vol.~58, no.~6, pp. 3014--3031, July 2009.

\bibitem{ieee80222}
\emph{Draft Standard for Wireless Regional Area Networks Part 22: Cognitive
  Wireless {RAN} Medium Access Control ({MAC}) and Physical Layer ({PHY})
  Specifications: Policies and Procedures for Operation in the {TV} Bands},
  IEEE Std. P802.22 Draft 7.0, Dec. 2010.

\bibitem{stevenson_commag09}
C.~R. {Stevenson}, G.~{Chouinard}, Z.~{Lei}, W.~{Hu}, S.~J. {Shellhammer}, and
  W.~{Caldwell}, ``{IEEE 802.22}: The first cognitive radio wireless regional
  area network standard,'' \emph{{IEEE} Commun. Mag.}, vol.~47, no.~1, pp.
  130--138, Jan. 2009.

\bibitem{jeon_twc_2008}
W.~S. {Jeon}, D.~G. {Jeong}, J.~A. {Han}, G.~{Ko}, and M.~S. {Song}, ``An
  efficient quiet period management scheme for cognitive radio systems,''
  \emph{{IEEE} Trans. Wireless Commun.}, vol.~7, no.~2, pp. 505--509, Feb.
  2008.

\bibitem{luo_twc_2009}
L.~{Luo}, N.~M. {Neihart}, S.~{Roy}, and D.~J. {Allstot}, ``A two-stage sensing
  technique for dynamic spectrum access,'' \emph{{IEEE} Trans. Wireless
  Commun.}, vol.~8, no.~6, pp. 3028--3037, June 2009.

\bibitem{Park_arxiv_2010}
\BIBentryALTinterwordspacing
J.~Park, P.~Pawe{\l}czak, P.~{Gr{\o}nsund}, and D.~\v{C}abri\'{c}. Subchannel
  notching and channel bonding: Comparative analysis of opportunistic spectrum
  {OFDMA} designs. [Online]. Available: \url{http://arxiv.org/abs/1007.5080}
\BIBentrySTDinterwordspacing

\bibitem{polydoros_tcom_1984}
A.~{Polydoros} and C.~L. {Weber}, ``A unified approach to serial search
  spread-spectrum code acquisition---part {I}: General theory,'' \emph{{IEEE}
  Trans. Commun.}, vol. COM-32, no.~5, pp. 542--549, May 1984.

\bibitem{suwansantisuk_milcom_2008}
W.~{Suwansantisuk} and M.~Z. {Win}, ``Ultrawide bandwidth serial-search
  multi-dwell acquisition---part {I}: Asymptotic theory,'' in \emph{Proc. IEEE
  MILCOM}, San Diego, CA, USA, Nov. 17--19, 2008.

\bibitem{suwansantisuk_izs_2006}
------, ``On the asymptotic performance of multi-dwell signal acquisition in
  dense multipath channels,'' in \emph{Proc. IEEE IZS}, Zurich, Switzerland,
  Feb. 22--24, 2006.

\bibitem{jeon_tvt_2010}
W.~S. {Jeon} and D.~G. {Jeong}, ``An advanced quiet-period management scheme
  for cognitive radio systems,'' \emph{{IEEE} Trans. Veh. Technol.}, vol.~59,
  no.~3, pp. 1242--1256, Mar. 2010.

\bibitem{vanmieghem_book_2006}
P.~{Van Mieghem}, \emph{Performance Analysis of Communications Networks and
  Systems}.\hskip 1em plus 0.5em minus 0.4em\relax Cambridge University Press,
  2006.

\bibitem{Park_arxiv_2009}
\BIBentryALTinterwordspacing
J.~Park, P.~Pawe{\l}czak, and D.~\v{C}abri\'{c}. (2010, Aug. 12) Performance of
  joint spectrum sensing and {MAC} algorithms for multichannel opportunistic
  spectrum access ad hoc networks. \emph{{IEEE} Trans. Mobile Comput.},
  accepted for publication. [Online]. Available:
  \url{http://arxiv.org/abs/0910.4704}
\BIBentrySTDinterwordspacing

\bibitem{Digham}
F.~F. {Digham}, M.-S. {Alouini}, and M.~K. {Simon}, ``On the energy detection
  of unknown signals over fading channels,'' in \emph{Proc. IEEE ICC},
  Anchorage, AK, USA, May 11--15, 2003.

\bibitem{wellens_phycom_2009}
M.~{Wellens}, J.~{Riihij{\"a}rvi}, and P.~{M{\"a}h{\"o}nen}, ``Empirical time
  and frequency domain models of spectrum use,'' \emph{Elsevier Physical
  Communication Journal}, vol.~2, no. 1--2, pp. 10--32, Mar.--Jun. 2009.

\end{thebibliography}
\end{document}